\documentclass[useAMS,usenatbib]{mnras}
\usepackage{graphicx}

\usepackage{journal_names}
\usepackage{subfig}
\usepackage{pdflscape}
\usepackage{amssymb}

\title[VLBI sources at z$>$4.5]{On the nature of bright compact radio sources at z$>$4.5} 
\author[Rocco Coppejans et al.]{Rocco Coppejans$^{1}$\thanks{E-mail: r.coppejans@astro.ru.nl }, S\'{a}ndor Frey$^{2}$, D\'{a}vid Cseh$^{1}$, Cornelia M\"{u}ller$^{1}$, Zsolt Paragi$^{3}$, \newauthor Heino Falcke$^{1,4}$, Krisztina \'{E}. Gab\'{a}nyi$^{2,5}$, Leonid I. Gurvits$^{3,6}$, Tao An$^{7,8}$ \newauthor and Oleg Titov$^{9}$ \\
$^{1}$Department of Astrophysics/IMAPP, Radboud University, P.O. Box 9010, 6500 GL Nijmegen, The Netherlands\\
$^{2}$F\"{O}MI Satellite Geodetic Observatory, PO Box 585, H-1592 Budapest, Hungary\\
$^{3}$Joint Institute for VLBI ERIC, Postbus 2, 7990 AA Dwingeloo, The Netherlands\\
$^{4}$Netherlands Institute for Radio Astronomy (ASTRON), PO Box 2, 7990 AA Dwingeloo, The Netherlands\\
$^{5}$Konkoly Observatory, MTA Research Centre for Astronomy and Earth Sciences, PO Box 67, H-1525 Budapest, Hungary\\
$^{6}$Department of Astrodynamics \& Space Missions, Delft University of Technology, 2629 HS Delft, The Netherlands\\
$^{7}$Shanghai Astronomical Observatory, Chinese Academy of Sciences, 80 Nandan Road, 200030 Shanghai, P. R. China\\
$^{8}$Key Laboratory of Radio Astronomy, Chinese Academy of Sciences, 210008 Nanjing, P. R. China\\
$^{9}$Geoscience Australia, PO Box 378, Canberra, ACT 2601, Australia\\
}

\begin{document}

\date{}

\pagerange{\pageref{firstpage}--\pageref{lastpage}} \pubyear{2016}

\maketitle

\label{firstpage}

\begin{abstract}
High-redshift radio-loud quasars are used to, among other things, test the predictions of cosmological models, set constraints on black hole growth in the early universe and understand galaxy evolution. Prior to this paper, 20 extragalactic radio sources at redshifts above 4.5 have been imaged with very long baseline interferometry (VLBI). Here we report on observations of an additional ten $z>4.5$ sources at 1.7 and 5\,GHz with the European VLBI Network (EVN), thereby increasing the number of imaged sources by 50\,per\,cent. Combining our newly observed sources with those from the literature, we create a substantial sample of 30 $z>4.5$ VLBI sources, allowing us to study the nature of these objects. Using spectral indices, variability and brightness temperatures, we conclude that of the 27 sources with sufficient information to classify, the radio emission from one source is from star formation, 13 are flat-spectrum radio quasars and 13 are steep-spectrum sources. We also argue that the steep-spectrum sources are off-axis (unbeamed) radio sources with rest-frame self-absorption peaks at or below GHz frequencies and that these sources can be classified as gigahertz peaked-spectrum (GPS) and megahertz peaked-spectrum (MPS) sources.
\end{abstract}


\section{Introduction}
\label{sec:introduction}
High-redshift quasars are among the most intriguing objects because they are thought to be associated with the youngest supermassive ($10^6$ to $10^9$\,M$_\odot$) black holes in the Universe. These accreting black holes play a key role in the evolution of their host galaxies via feedback \citep[e.g.][]{2005MNRAS.362...25B,fabian2012,2013Sci...341.1082M}. The observed properties of the highest-redshift black holes set constraints on their accretion process and thus on the black hole growth \citep[e.g.][]{2012MNRAS.425.2892W,2014MNRAS.440L..91P}. They are also indispensable for 21-cm absorption studies as they serve as illuminating background objects \citep[e.g.][]{2012RPPh...75h6901P}. However, their evolution and radio loudness are not well understood. Whether these properties are related to the accretion process or to the density of the cosmic microwave background (CMB) photons at these redshifts is still an open question \citep{2014MNRAS.442L..81F}.

The compact core--jet structures in high-redshift radio-loud quasars are valuable additions to the samples used for classical cosmological tests like the apparent angular size--redshift \citep[e.g.][]{1999A&A...342..378G} and the apparent proper motion--redshift relations \citep[e.g.][]{1999NewAR..43..757K}. The redshift $z=4.5$ corresponds to less than 10\,per\,cent of the present age of the Universe. Here the predictions of cosmological models can be radically different, yet useful test objects are sparse. These tests require high-resolution radio interferometric data on the compact radio structures and their variation with time.

High-redshift radio-loud quasars provide critical input into source counts and quasar luminosity function studies \citep[e.g.][]{2004ApJ...612..698H}. From geometrical considerations of the jet inclination angles with respect to the line of sight, we expect that for each active galactic nucleus (AGN) whose jet is pointed within a small angle of our line of sight \citep[blazars; e.g.][]{1999ASPC..159....3U,2013FrPhy...8..609K}, there should be hundreds\footnote{\citet{2016arXiv160305684G} showed that for every blazar that we observe with a viewing angle $\theta<1/\Gamma$ (where $\Gamma$ is the Lorentz factor), there exists $2\Gamma^2$ sources with $\theta>1/\Gamma$.} of sources with jets pointing elsewhere \citep{2011MNRAS.416..216V}. In \citet{2011MNRAS.416..216V}, the authors compare the number of high-redshift radio-loud sources in the Sloan Digital Sky Survey (SDSS) and Very Large Array (VLA) Faint Images of the Radio Sky at Twenty-centimeter (FIRST) survey \citep{FIRST} with the number of blazars between $z=1$ and $z=6$ and find that they are consistent at $z<3$ but disagree at $z>3$. Beyond $z=3$, the number of high-redshift radio-loud sources is significantly lower than what is expected from the number of blazars at these redshifts. The authors propose three possible explanations for the discrepancy: 1) the average bulk Lorentz factor decreases as a function of redshift; 2) there is a bias in SDSS and FIRST against detecting high-redshift radio-loud sources; 3) there is a bias in SDSS against detecting high-redshift radio-loud and radio-quiet sources. The apparent lack of AGN with misaligned jets at very hight redshifts led \citet{2016arXiv160305684G} to propose a model in which a dusty ``bubble'' surrounding the central regions obscures the nucleus. The AGN is visible in the optical only if it is observed along the jet which cleared up the obscuring material in that direction.

The reason for the missing misaligned high-redshift sources can be investigated using very long baseline interferometry (VLBI) observations of the know high-redshift sources. By determining the variability properties, spectral indices, Doppler boosting and morphologies of the sources from VLBI observations, the sources can be classified as blazars or misaligned sources. In the case of resolved sources, the VLBI observations can be used as a first epoch to measure the jet proper motion from which the Lorentz factor can be calculated \citep[e.g.][]{Frey2015}.

In this paper, we present 1.7 and 5\,GHz VLBI observations of ten $z>4.5$ sources conducted with the European VLBI Network (EVN). These observations increase the number of $z>4.5$ sources that have been imaged with VLBI from 20 to 30. Combining our new observations with those from the literature, we investigate the nature of the $z>4.5$ VLBI sources. In Section \ref{sec:obsrve and reduce}, we describe how we selected the sources and reduced the EVN data. The source properties derived from the images are presented in Section \ref{sec:results}. Section \ref{sec:vlbi_z>4.5} contains a summary of the properties of the $z>4.5$ sources that have previously been imaged with VLBI. In Section \ref{sec:discus} we discuss the origin of the radio emission, variability properties, spectral indices and Doppler boosting of the sources, which we then use to classify them in Section \ref{sec:what are they?}. Finally, a summary and conclusion are presented in Section \ref{sec:summary}. The following cosmological model parameters are assumed throughout this paper: $\Omega_{\rm m}=0.3$, $\Omega_{\lambda}=0.7$, $H_0=72$\,km\,s$^{-1}$\,Mpc$^{-1}$.

\section{Target selection, Observations, data reduction and VLBI images}
\label{sec:obsrve and reduce}

\subsection{Target selection}
\label{subsec:target slect}
The radio-emitting sources with spectroscopic redshifts greater than 4.5 in the SDSS Data Release 10 (DR10) quasar catalogue \citep{2014A&A...563A..54P} and the Million quasars catalogue \citep{2015PASA...32...10F}\footnote{http://quasars.org/milliquas.htm} were considered for observation. From this list we selected all of the sources that are unresolved ($<5$\,arcsec) in the FIRST survey and have integrated 1.4\,GHz flux densities exceeding $5$\,mJy. In addition to the catalogue samples above, the spectroscopic redshifts of J1013+2811, J1454+1109 and J1628+1154 have recently been measured by \citet{2013AJ....146...10T}. The final list of $z>4.5$ sources we observed with the EVN is given in Table~\ref{tbl:image parameters}.

\subsection{Observing setup}
\label{subsec:Observing setup}
The sources were observed with the EVN at central frequencies of 1.658 and 4.990\,GHz during six project segments: EC052A, EC052B, EC052C, EC052D, EC052E and EC052F. In Table~\ref{tbl:obsrvations details}, the observing frequency, the date of the observations and the radio telescopes that successfully participated in each project segment are shown. The following radio observatories participated in the experiments: Effelsberg (Ef; Germany), Hartebeesthoek (Hh; South Africa), Jodrell Bank Mk2 (Jb; United Kingdom), Onsala (On; Sweden), Toru\'{n} (Tr; Poland), Noto (Nt; Italy), Medicina (Mc; Italy), Sheshan (Sh; China), the Westerbork Synthesis Radio Telescope (Wb; the Netherlands) and Yebes (Ys; Spain). 

The observations at both frequencies were done using 2\,s integrations at a data rate of 1024\,Mbit\,$\mathrm{s^{-1}}$ in left and right circular polarizations with eight subbands per polarization and 16\,MHz of bandwidth per subband. The technique of electronic VLBI\footnote{ See http://www.jive.eu/e-vlbi and www.evlbi.org/evlbi } was used, where the data are streamed to the central correlator using optical fiber networks in real time. The observations of each target source were interleaved with observations of a phase calibrator. The same phase calibrator (listed in Table~\ref{tbl:phase calibrator}) was used for each target source at both observing frequencies. The distances between the target sources and their phase calibrators are given in Table~\ref{tbl:phase calibrator}. Additionally, VLBI images and flux densities at two or more frequencies between 2.3 and 8.3\,GHz of all of the phase calibrators are available in the Astrogeo data base\footnote{http://astrogeo.org/, maintained by L. Petrov.}. In Table~\ref{tbl:image parameters}, the project segments during which each source were observed at each frequency are given. 

Three radio telescopes experienced problems during EC052A resulting in only the six stations shown in Table~\ref{tbl:obsrvations details} successfully producing data. The sources that were initially observed at 1.7\,GHz during EC052A were therefore re-observed during EC052C, EC052E and EC052F. This resulted in some of the sources being observed two or three times at 1.7\,GHz.

\begin{table*}
 \hspace{-4cm}
 \centering
 \begin{minipage}{\columnwidth}
  \caption{Details of the observations}
  \begin{tabular}{cccc}
  \hline
  Project segment & $\nu$ [GHz] & Observing date & Participating radio telescopes \\
  \hline  
  EC052A & 1.7 & 2014 Oct 14 & Ef, Hh, Jb, On, Tr, Sh\\
  EC052B & 5.0   & 2014 Nov 18 & Ef, Hh, Jb, On, Tr, Nt, Ys, Wb, Sh\\
  EC052C & 1.7 & 2015 Feb 10 & Ef, Hh, Jb, On, Tr, Mc, Wb, Sh\\
  EC052D & 5.0   & 2015 Mar 24 & Ef, Hh, Jb, On, Tr, Nt, Ys, Sh\\
  EC052E & 1.7 & 2015 Jun 23 & Ef, Hh, Jb, On, Tr, Mc, Wb, Sh\\ 
  EC052F & 1.7 & 2016 Jan 12 & Ef, Hh, Jb, On, Tr, Mc, Wb\\ 
  \hline
  \end{tabular}
  \label{tbl:obsrvations details}
 \end{minipage}
\end{table*}

\begin{table*}
 \hspace{-6cm}
 \begin{minipage}{\columnwidth}
  \caption{Details of the phase calibrators}
  \begin{tabular}{ccccc}
  \hline
  Phase      & Source      & Separation   & Unresolved 1.7\,GHz flux  & Unresolved 5\,GHz flux \\
  calibrator & ID  & [$^{\circ}$] & density [mJy]  & density [mJy] \\
  \hline  
  J0010+1724   & J0011+1446   &  2.9 & 200 & 100  \\
  J0216$-$0105 & J0210$-$0018 &  1.6 & 50  & 60  \\
  J0950+0615   & J0940+0526   &  2.6 & 60  & 60  \\
  J1023+2856   & J1013+2811   &  2.3 & 40  & 50  \\
  J1321+2216   & J1311+2227   &  2.3 & 60  & 170  \\
  J1350+3034   & J1400+3149   &  2.4 & 150 & 170  \\
  J1453+1036   & J1454+1109   &  0.7 & 100 & 50  \\
  J1544+3240   & J1548+3335   &  1.3 & 50  & 70  \\  
  J1622+1426   & J1628+1154   &  2.9 & 50  & 50  \\
  J1726+3213   & J1720+3104   &  1.7 & 70  & 150  \\
  \hline
  \multicolumn{5}{p{12cm}}{\footnotesize{\textbf{Columns:} Col.~1 -- phase calibrator name (J2000); Col.~2 -- target source name (J2000); Col.~3 -- angular separation between the target source and the phase calibrator; Col.~4 -- approximate 1.7\,GHz correlated flux density at the longest baselines; Col.~5 -- approximate 5\,GHz correlated flux density at the longest baselines.}}\\
  \end{tabular}
  \label{tbl:phase calibrator}
 \end{minipage}
\end{table*}

\begin{table*}
 \hspace{-6cm}
 \begin{minipage}{\columnwidth}
  \caption{Image parameters}
  \begin{tabular}{ccccccc}
  \hline
  ID & FIRST flux     & $\nu$ & Project segment & \multicolumn{2}{c}{Restoring beam} & $1\sigma$ image noise\\
     & density [mJy]  & [GHz] &              & [mas$\times$mas] & PA [$^{\circ}$] & [mJy\,beam$^{-1}$]\\
  \hline  
  J0011+1446   & $24.3\pm1.2$ & 1.7 & EC052A, C  & $3.0\times6.6$ & 82.3 & 0.11\\
               &            & 5.0   & EC052B     & $1.4\times1.8$ & 64.4 & 0.14\\
  \hline
  J0210$-$0018 & $8.5\pm0.4$ & 1.7 & EC052A, C  & $3.1\times6.6$ & 80.2 & 0.10\\
               &            & 5.0   & EC052B     & $1.5\times1.8$ & 59.6 & 0.06\\
  \hline
  J0940+0526   & $58.5\pm2.9$ & 1.7 & EC052A, E, F & $3.2\times6.1$ & 80.8 & 0.17\\
               &            & 5.0   & EC052B     & $1.4\times1.5$ & 30.9 & 0.12\\
  \hline
  J1013+2811   & $14.4\pm0.7$ & 1.7 & EC052A, C, F & $3.3\times7.0$ & 82.9 & 0.06\\
               &            & 5.0   & EC052B     & $1.5\times1.6$ & 47.9 & 0.10\\
  \hline
  J1311+2227   & $6.5\pm0.3$ & 1.7 & EC052A, C   & $3.1\times6.0$ & 79.8 & 0.05\\
               &            & 5.0   & EC052B     & $1.4\times1.6$ & 35.7 & 0.03\\
  \hline
  J1400+3149   & $20.5\pm1.0$ & 1.7 & EC052E     & $4.1\times5.9$ & 80.8 & 0.06\\
               &            & 5.0   & EC052D     & $1.3\times1.6$ & 28.0 & 0.15\\
  \hline
  J1454+1109   & $15.1\pm0.8$ & 1.7 & EC052E     & $3.5\times5.5$ & 86.4 & 0.08\\
               &            & 5.0   & EC052D     & $1.4\times1.5$ & 45.2 & 0.10\\
  \hline
  J1548+3335   & $37.8\pm1.9$ & 1.7 & EC052A, F   & $3.4\times9.4$ & 83.9 & $0.06\,\&\,0.09^{\mathrm a}$\\  
               &            & 5.0   & EC052B     & $1.6\times1.8$ & 52.9 & 0.06\\
  \hline
  J1628+1154   & $41.0\pm2.0$ & 1.7 & EC052A, E   & --- & --- & ---\\
               &            & 5.0   & EC052B     & $1.5\times1.7$ & 63.0 & 0.06\\
  \hline
  J1720+3104   & $10.6\pm0.5$ & 1.7 & EC052A, E   & $3.6\times5.7$ & 80.5 & 0.10\\
               &            & 5.0   & EC052B     & $1.5\times1.8$ & 36.8 & 0.15\\
  \hline
  \multicolumn{7}{p{13cm}}{\footnotesize{\textbf{Columns:} Col.~1 -- source name (J2000); Col.~2 -- the source flux density in the FIRST survey; Col.~3 -- observing frequency; Col.~4 -- project segment(s) during which the source was observed; Col.~5 -- Gaussian restoring beam size (FWHM) in Fig.~\ref{fig:VLBI images}; Col.~6 -- Gaussian restoring beam major axis position angle (measured from north through east) in Fig.~\ref{fig:VLBI images}; Col.~7 -- image noise in Fig.~\ref{fig:VLBI images}.}}\\
  \multicolumn{7}{p{13cm}}{\footnotesize{\textbf{Note:} $^{\mathrm a}$ The first noise value is for J1548+3335a and the second is for J1548+3335b.}}\\
  \end{tabular}
  \label{tbl:image parameters}
 \end{minipage}
\end{table*}

\subsection{Data reduction and VLBI images}
\label{subsec:Data reduction and VLBI images}
The EVN data were reduced using the NRAO Astronomical Image Processing System \citep[\textsc{aips},][]{2003ASSL..285..109G} software package. The visibility amplitudes were first calibrated using antenna gains and system temperatures measured at the radio telescopes. Next the phase calibrators were fringe-fitted before exporting their visibilities from \textsc{aips} for imaging in the Caltech \textsc{difmap} package \citep{difmap}. The imaging was done using several iterations of \textsc{clean} and phase self-calibration before doing a single round of amplitude self-calibration across the entire observing time as the solution interval using the \textsc{difmap} task \textsc{gscale}. This gave an antenna gain correction from each phase calibrator for each radio telescope. For each project segment, we calculated the median gain correction factors for each of the radio telescopes, which were applied to the visibility amplitudes of the phase calibrators and target sources in \textsc{aips}. 

The \textsc{difmap} \textsc{clean} component models of the phase calibrators were then used to improve the phase solutions of the phase calibrators in \textsc{aips} during fringe-fitting. The improved solutions of each phase calibrator were then applied to its target source before exporting the visibility data from \textsc{aips} for imaging in \textsc{difmap}. The naturally weighted images shown in Fig.~\ref{fig:VLBI images} were made using several rounds of \textsc{clean}. The image parameters are presented in Table~\ref{tbl:image parameters}. Phase-only self-calibration was only applied to the sources in which the sum of the \textsc{clean} component flux densities exceeded $\sim10$\,mJy. 

For the sources that where observed more than once at 1.7\,GHz during different project segments, the visibility data from all of the project segments were combined into a single data set. This is discussed in detail in Section~\ref{sec:results}. In Fig.~\ref{fig:VLBI images}, the images made from the combined visibility data sets are shown for these sources. The source J1548+3335 has two widely-separated components at 1.7\,GHz. For clarity, apart from the full view, zoomed-in images of both components are also shown in Fig.~\ref{fig:VLBI images}. 

Finally, using the visibility data we estimated the flux densities of the phase calibrators on the longest baselines for each project segment (Table~\ref{tbl:phase calibrator}). We note that from the Astrogeo data base, the phase calibrators show significant flux density variability. For the phase calibrators that were observed more than once at 1.7\,GHz, the value in Table~\ref{tbl:phase calibrator} is the minimum flux density.

\begin{figure*}
  \subfloat{
  \begin{minipage}{75mm}
    \centering
    \includegraphics[width=\columnwidth]{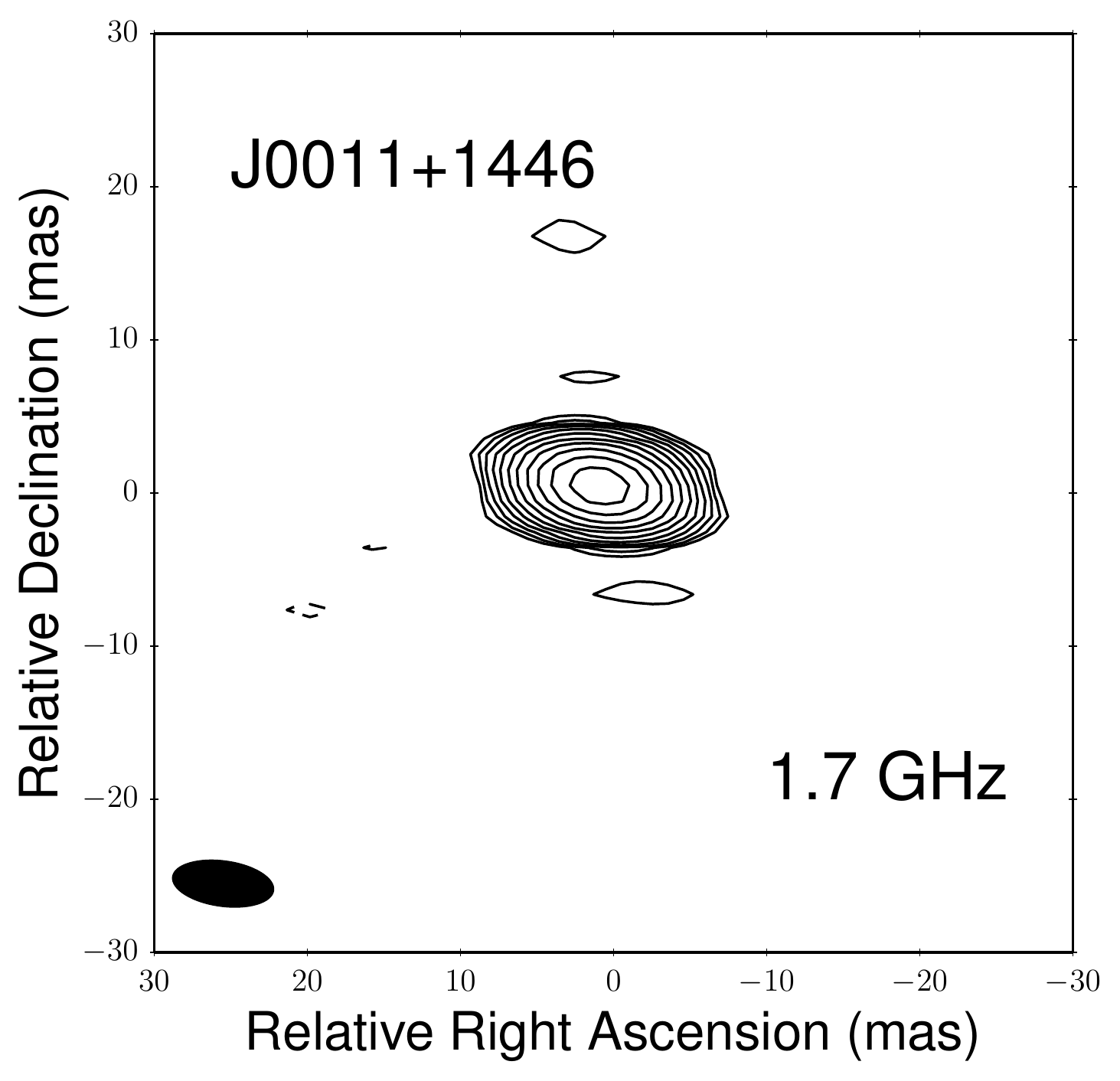}
  \end{minipage}  
  \begin{minipage}{72.5mm}
    \centering
    \includegraphics[width=\columnwidth]{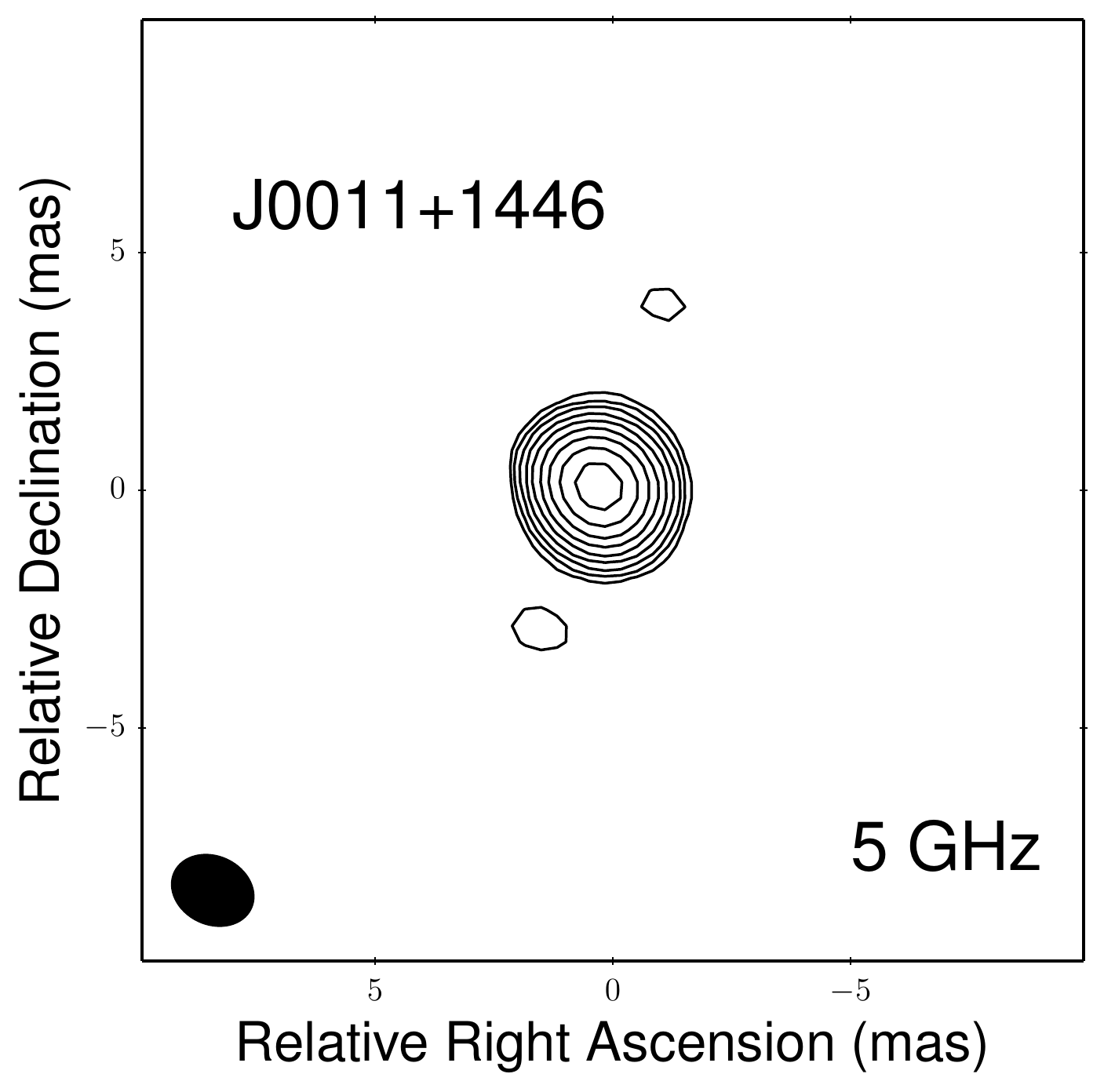}
  \end{minipage}}\\[-2ex] 
  \vspace{0.1cm}
  \subfloat{
  \begin{minipage}{75mm}
    \centering
    \includegraphics[width=\columnwidth]{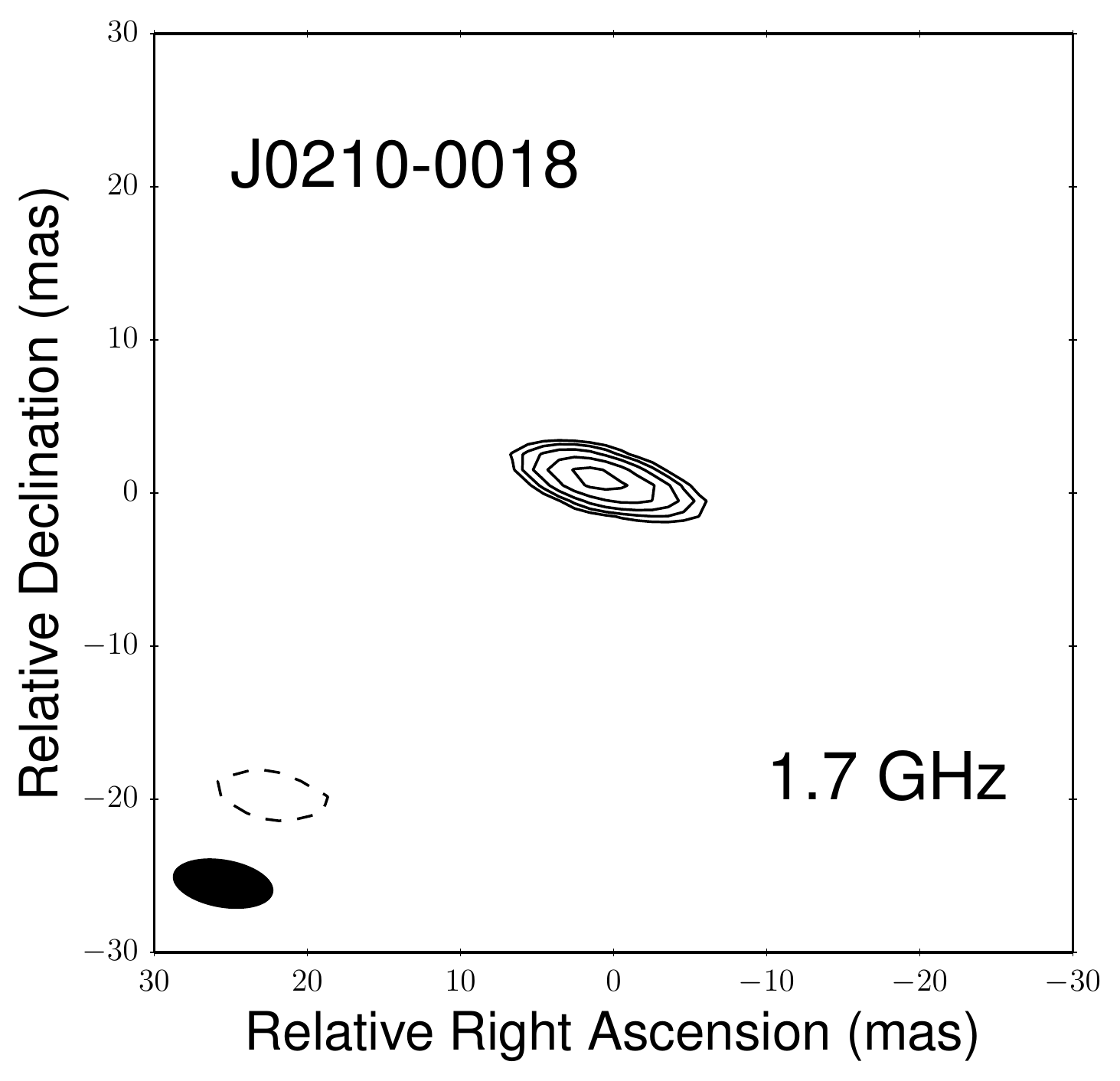}
  \end{minipage}  
  \begin{minipage}{72.5mm}
    \centering
    \includegraphics[width=\columnwidth]{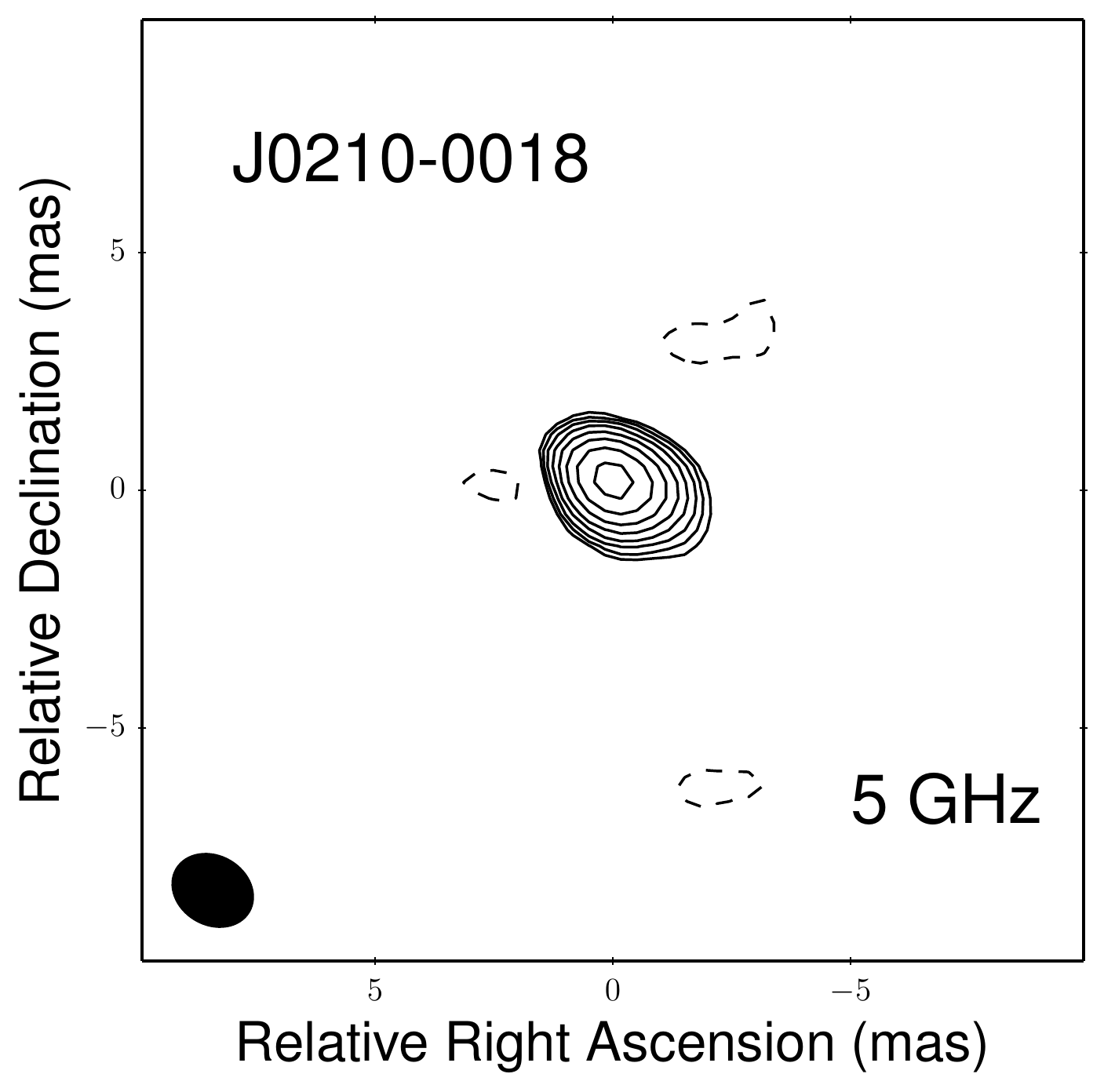}
  \end{minipage}}\\[-2ex]
  \vspace{0.1cm}
  \subfloat{
  \begin{minipage}{75mm}
    \centering
    \includegraphics[width=\columnwidth]{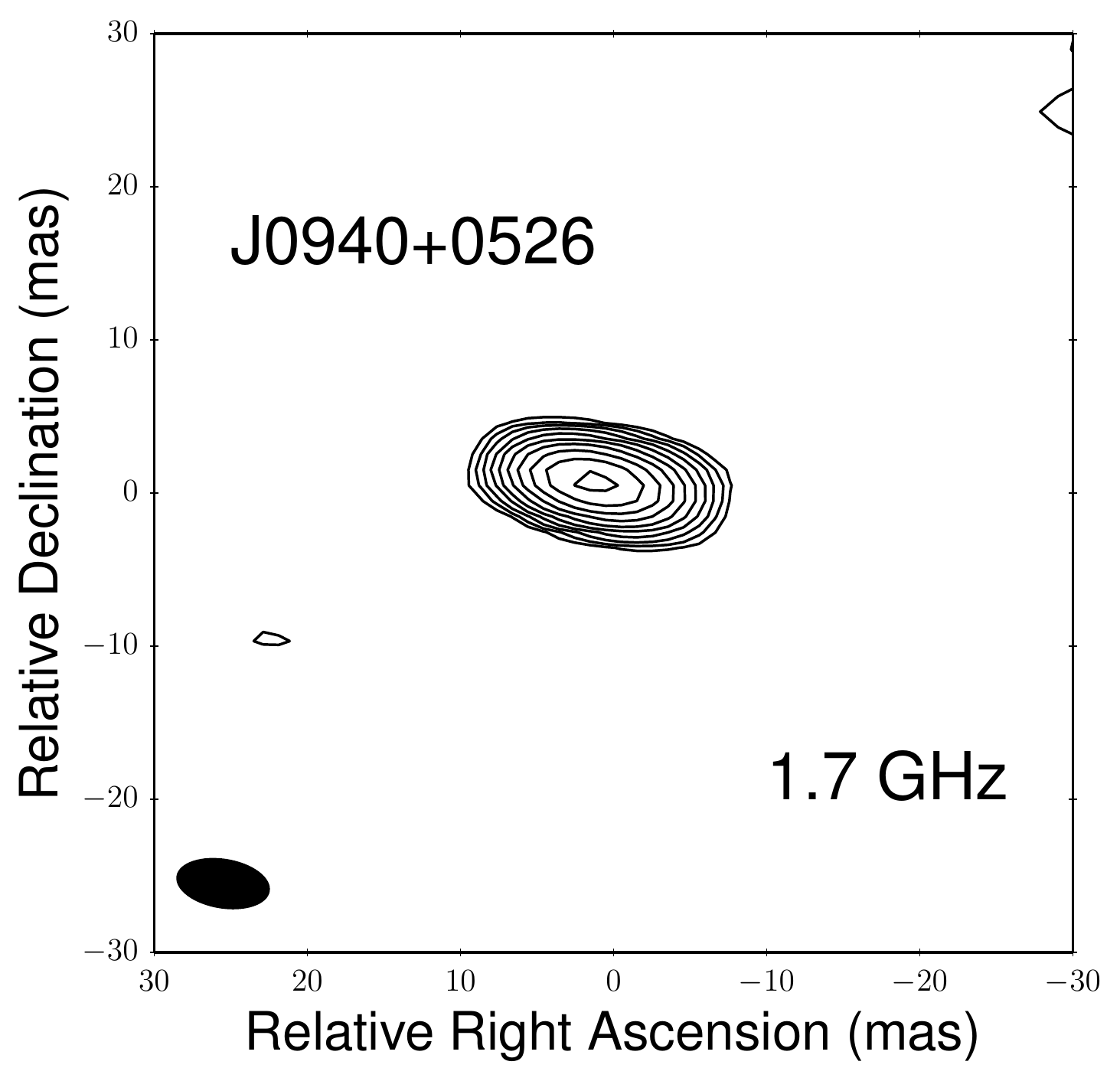}
  \end{minipage}  
  \begin{minipage}{72.5mm}
    \centering
    \includegraphics[width=\columnwidth]{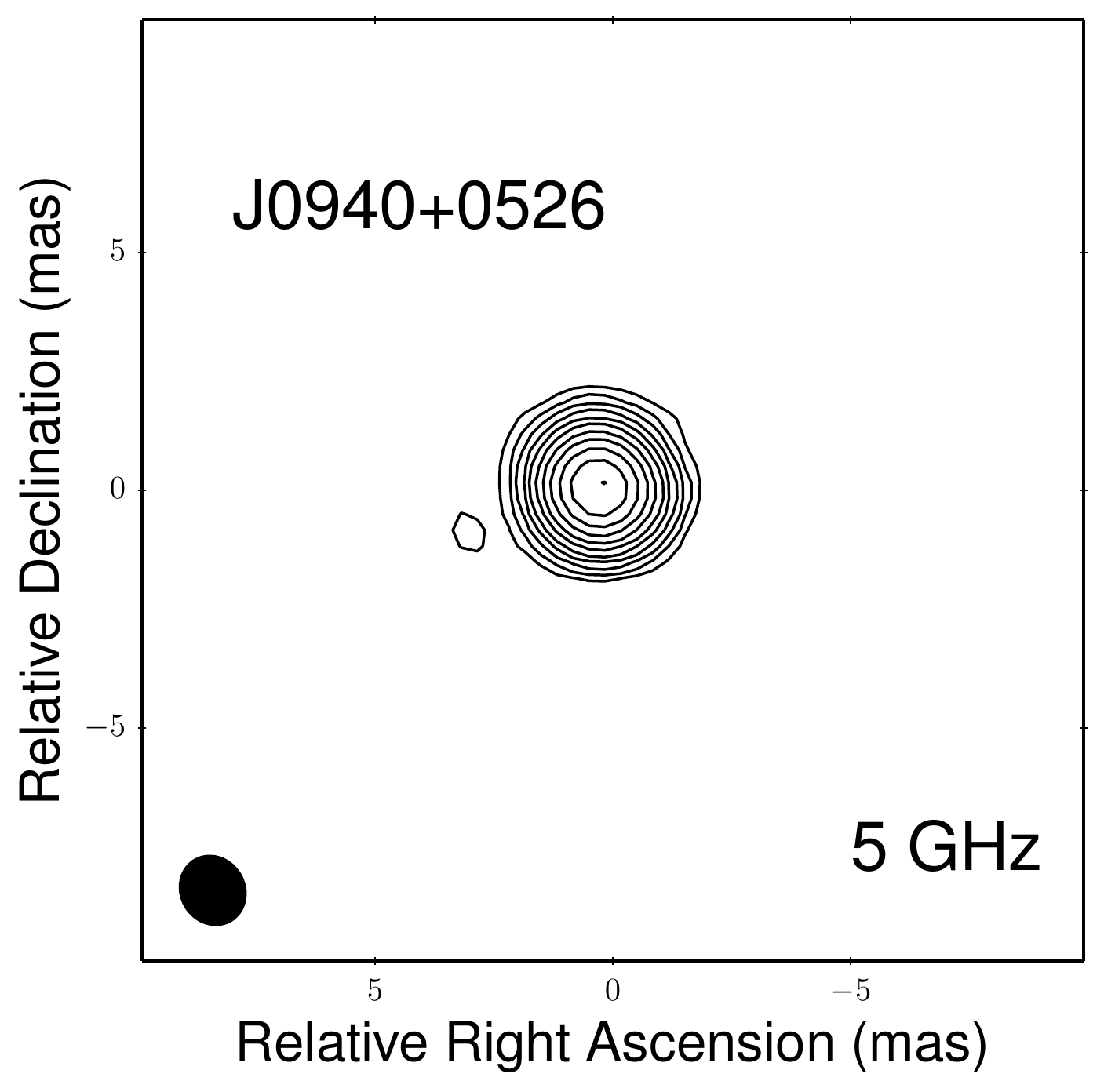}
  \end{minipage}}\\[-2ex]
  \caption{Naturally weighted 1.7 and 5\,GHz EVN images. The restoring beam (FWHM) is shown in the bottom left corner. The lowest contours are drawn at $-3$ and 3 times the image noise, the positive contours increase in factors of $\sqrt 2$ thereafter. For the source J1548+3335, separate images of its two 1.7\,GHz components are shown, as well as a zoomed-out version displaying both components in the same image.}
  \label{fig:VLBI images}
\end{figure*}

\begin{figure*}
  \subfloat{
  \begin{minipage}{75mm}
    \centering
    \includegraphics[width=\columnwidth]{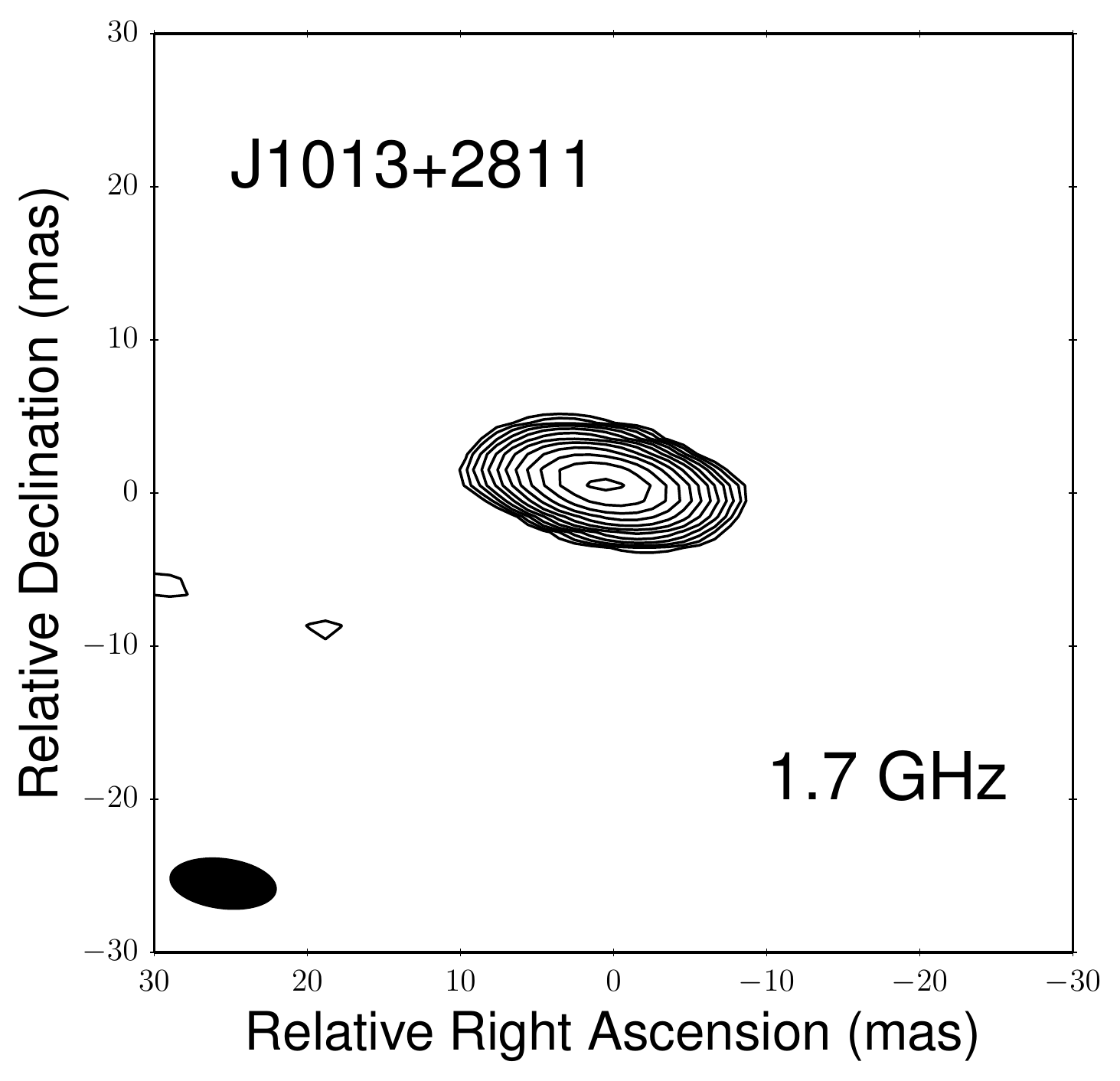}
  \end{minipage}  
  \begin{minipage}{72.5mm}
    \centering
    \includegraphics[width=\columnwidth]{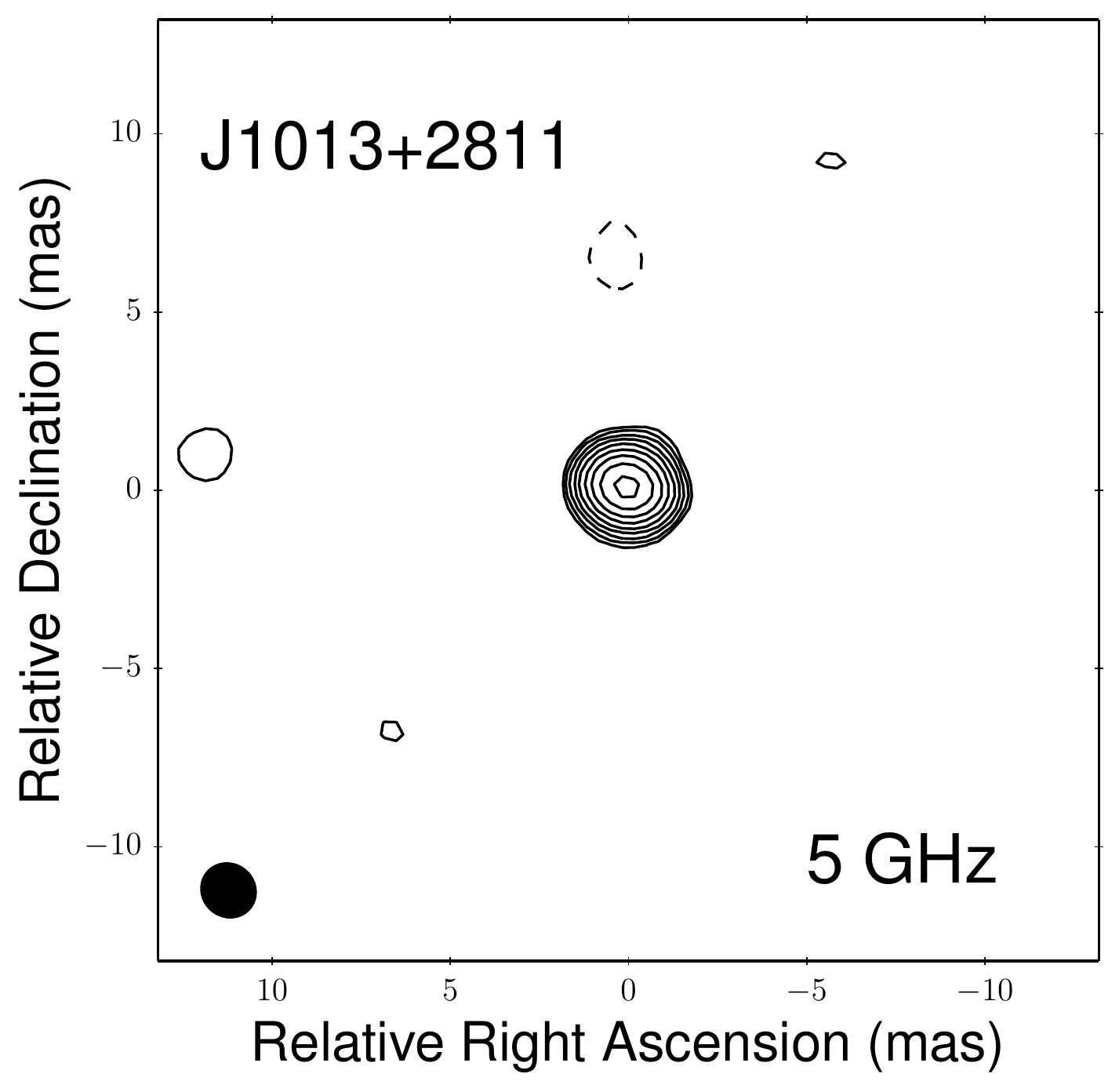}
  \end{minipage}}\\[-2ex]
  \vspace{0.1cm}
  \subfloat{
  \begin{minipage}{75mm}
    \centering
    \includegraphics[width=\columnwidth]{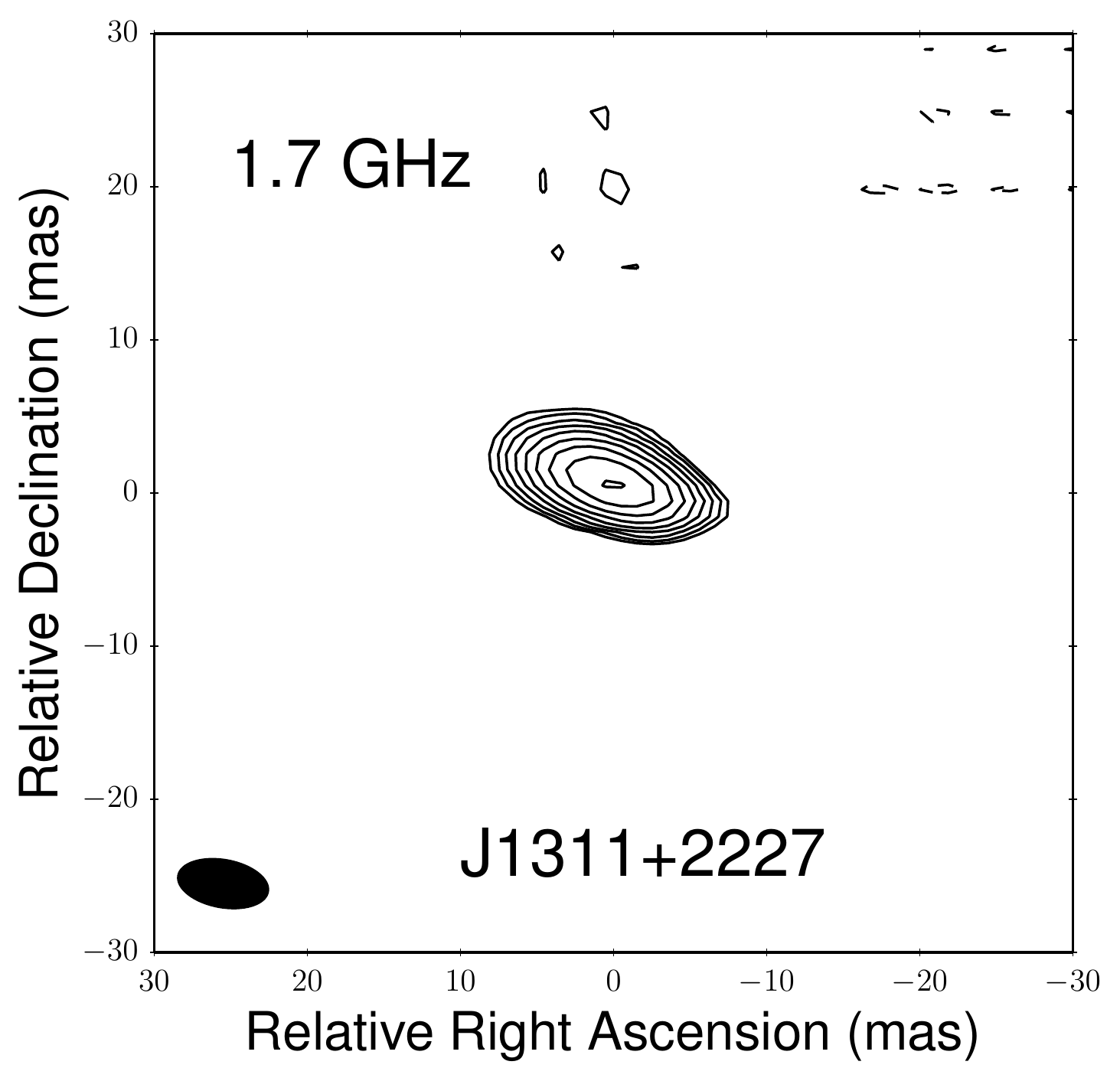}
  \end{minipage}  
  \begin{minipage}{72.5mm}
    \centering
    \includegraphics[width=\columnwidth]{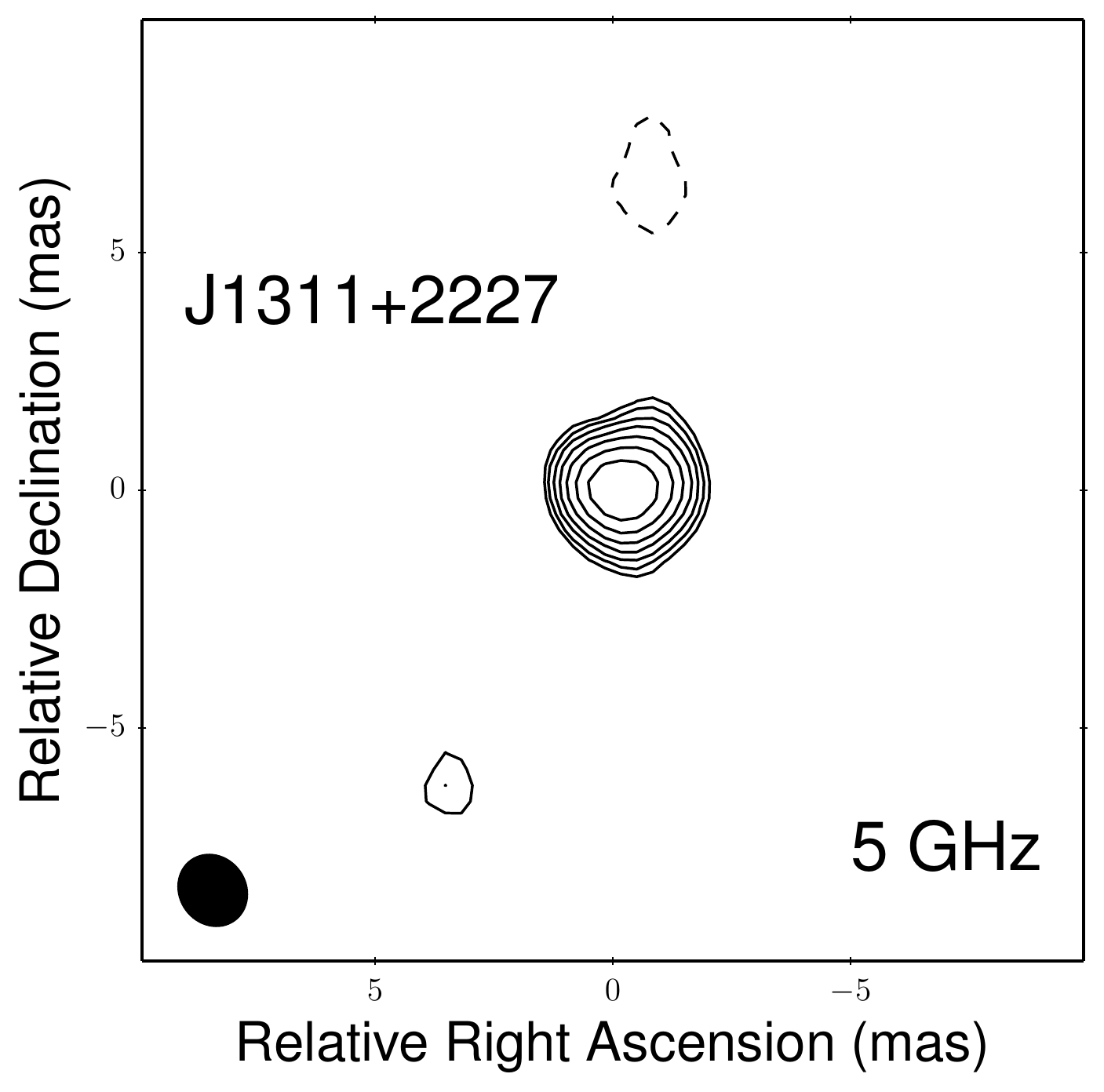}
  \end{minipage}}\\[-2ex]
  \vspace{0.1cm}
  \subfloat{
  \begin{minipage}{75mm}
    \centering
    \includegraphics[width=\columnwidth]{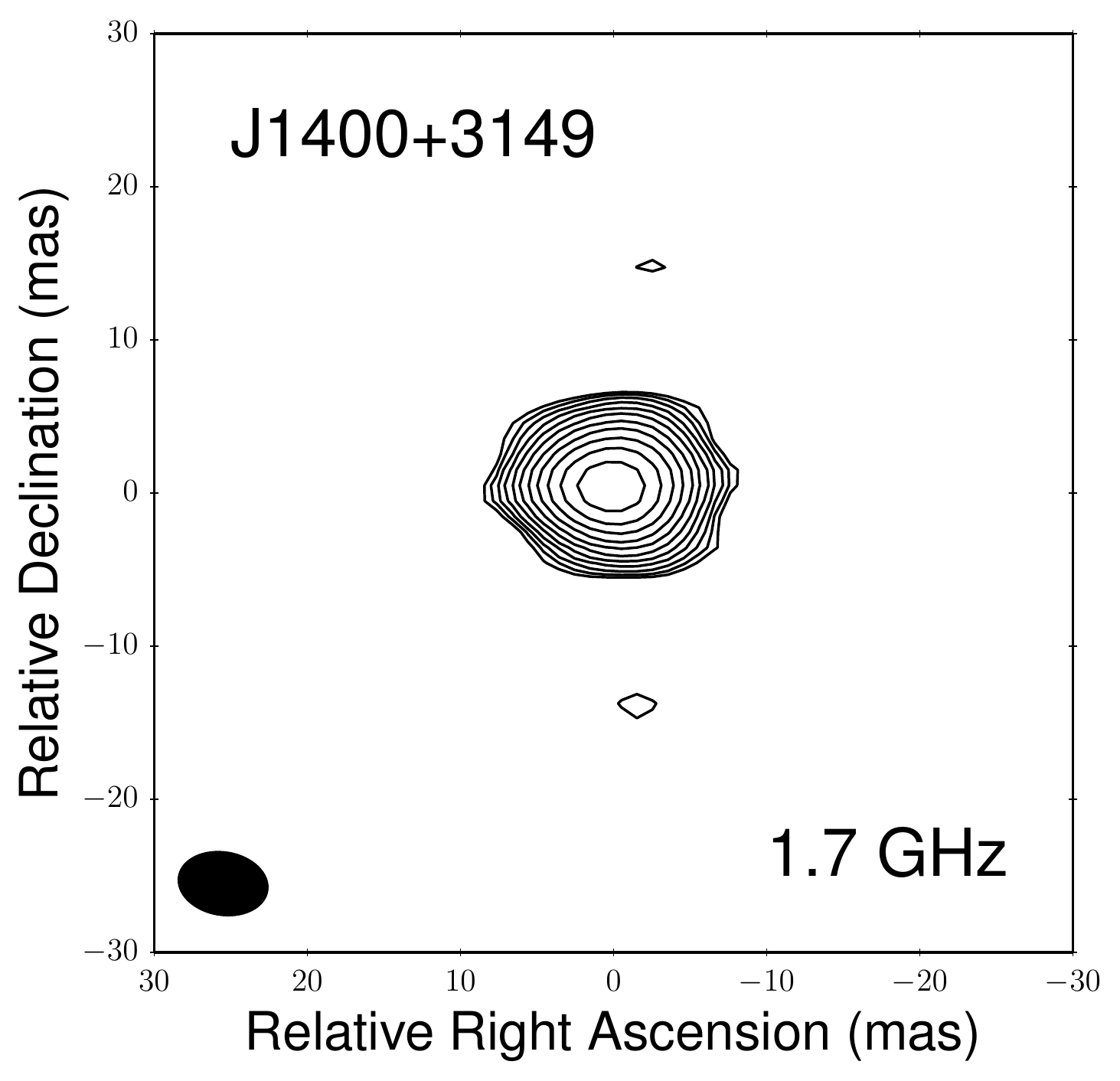}
  \end{minipage}  
  \begin{minipage}{72.5mm}
    \centering
    \includegraphics[width=\columnwidth]{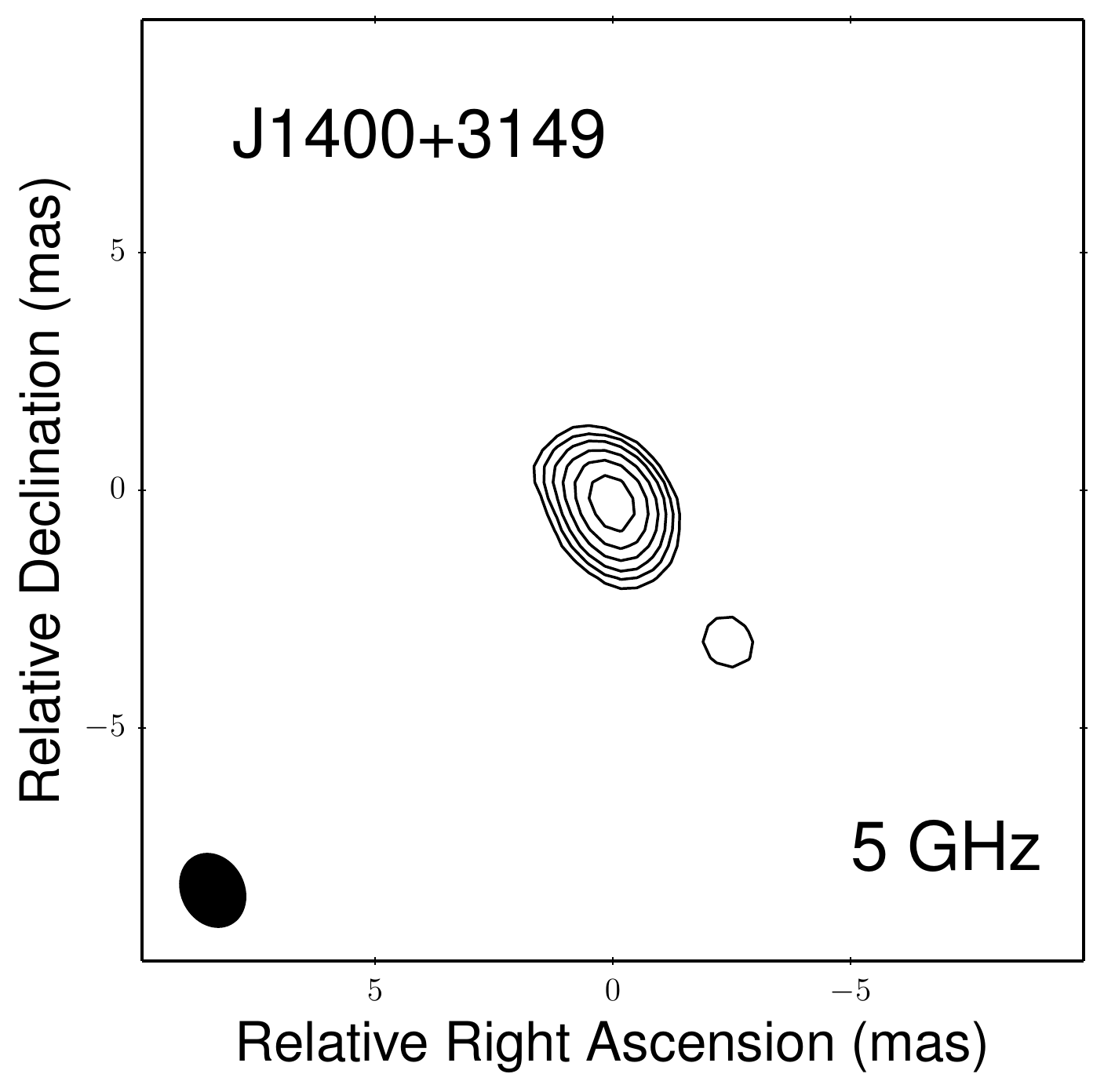}
  \end{minipage}}\\[-2ex]
  \contcaption{}
\end{figure*}

\begin{figure*}
  \subfloat{
  \begin{minipage}{75mm}
    \centering
    \includegraphics[width=\columnwidth]{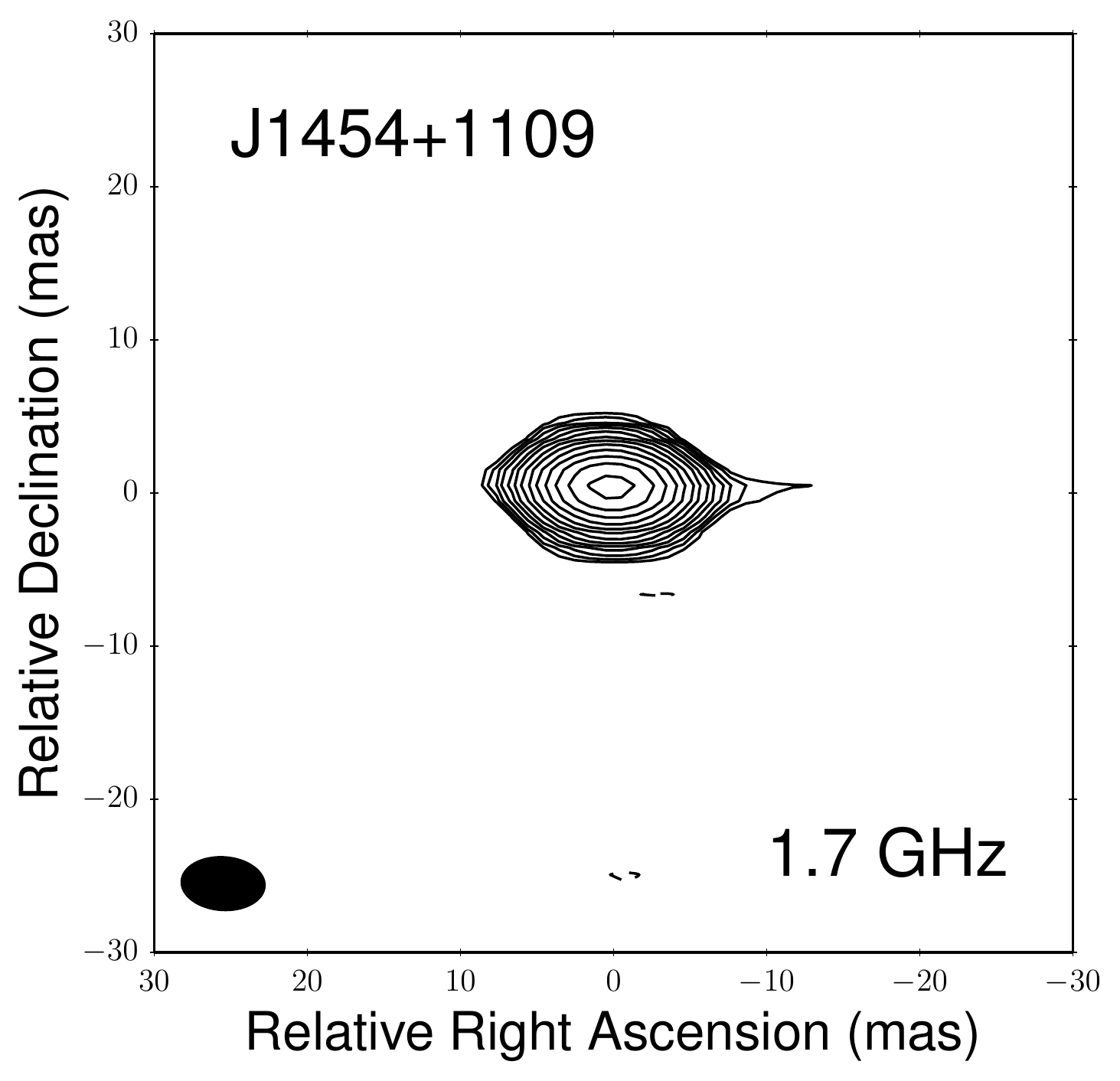}
  \end{minipage}  
  \begin{minipage}{72.5mm}
    \centering
    \includegraphics[width=\columnwidth]{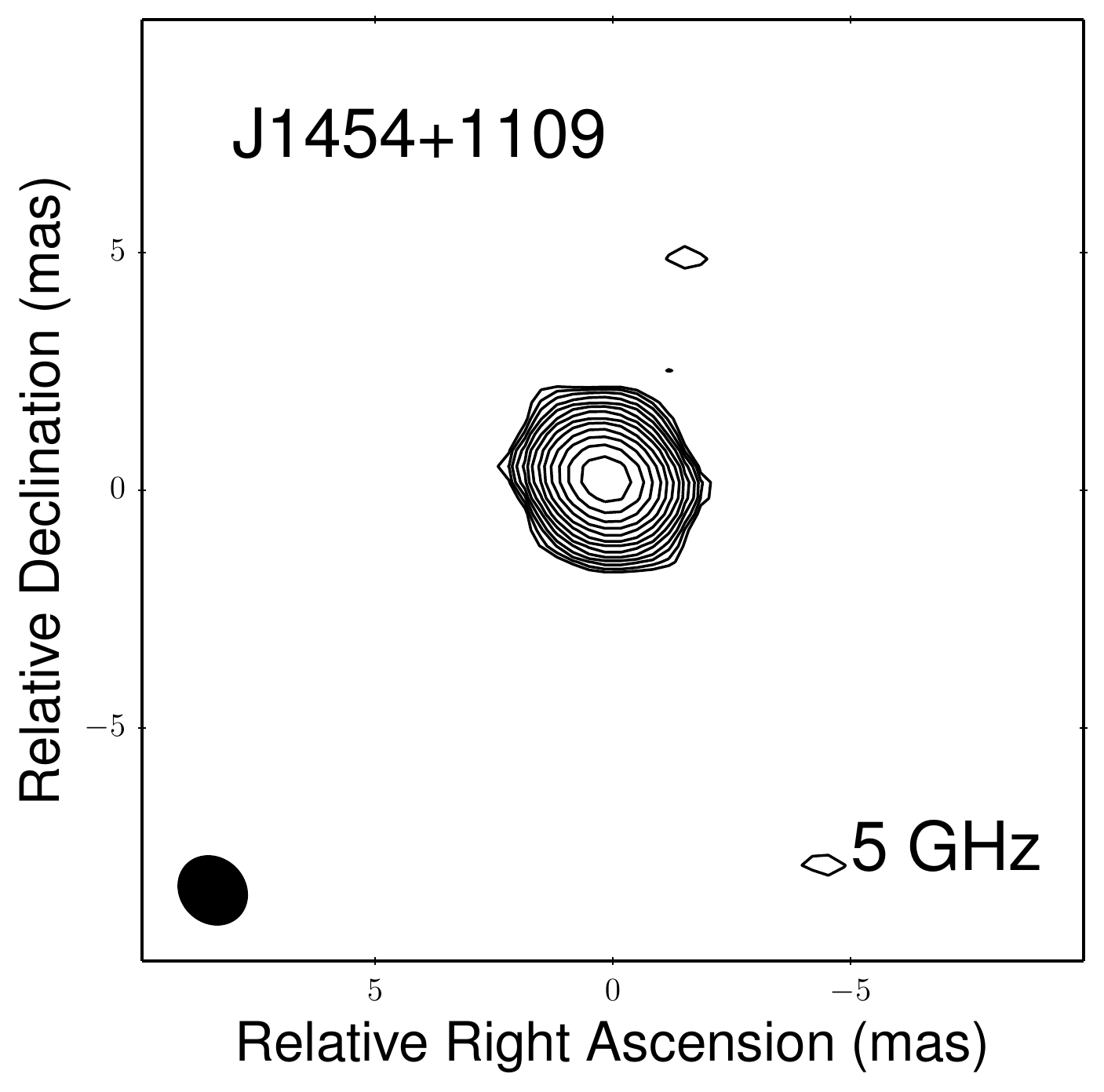}
  \end{minipage}}\\[-2ex]
  \vspace{0.1cm}
  \subfloat{
  \begin{minipage}{75mm}
    \centering
    \includegraphics[width=\columnwidth]{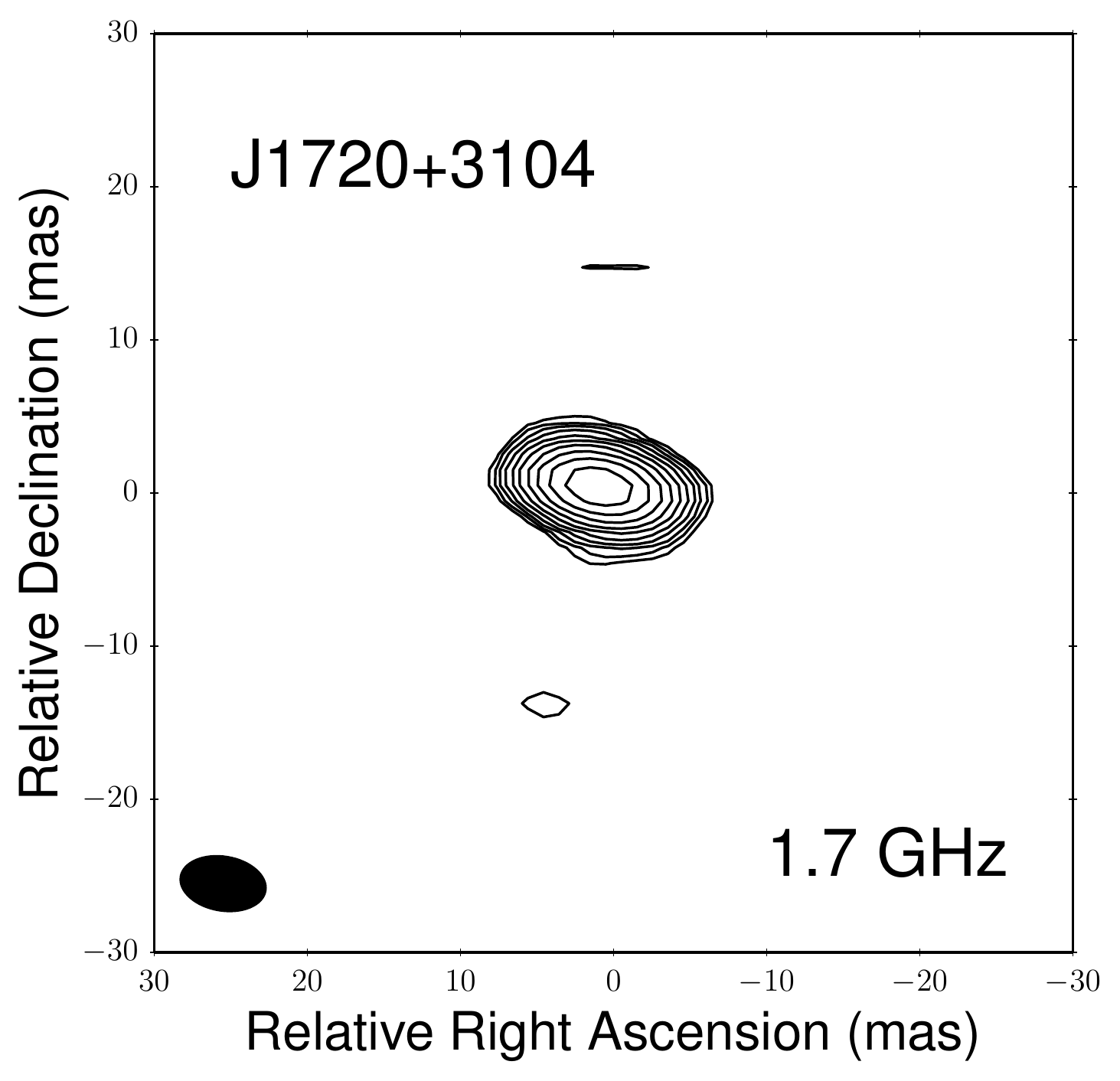}
  \end{minipage}  
  \begin{minipage}{72.5mm}
    \centering
    \includegraphics[width=\columnwidth]{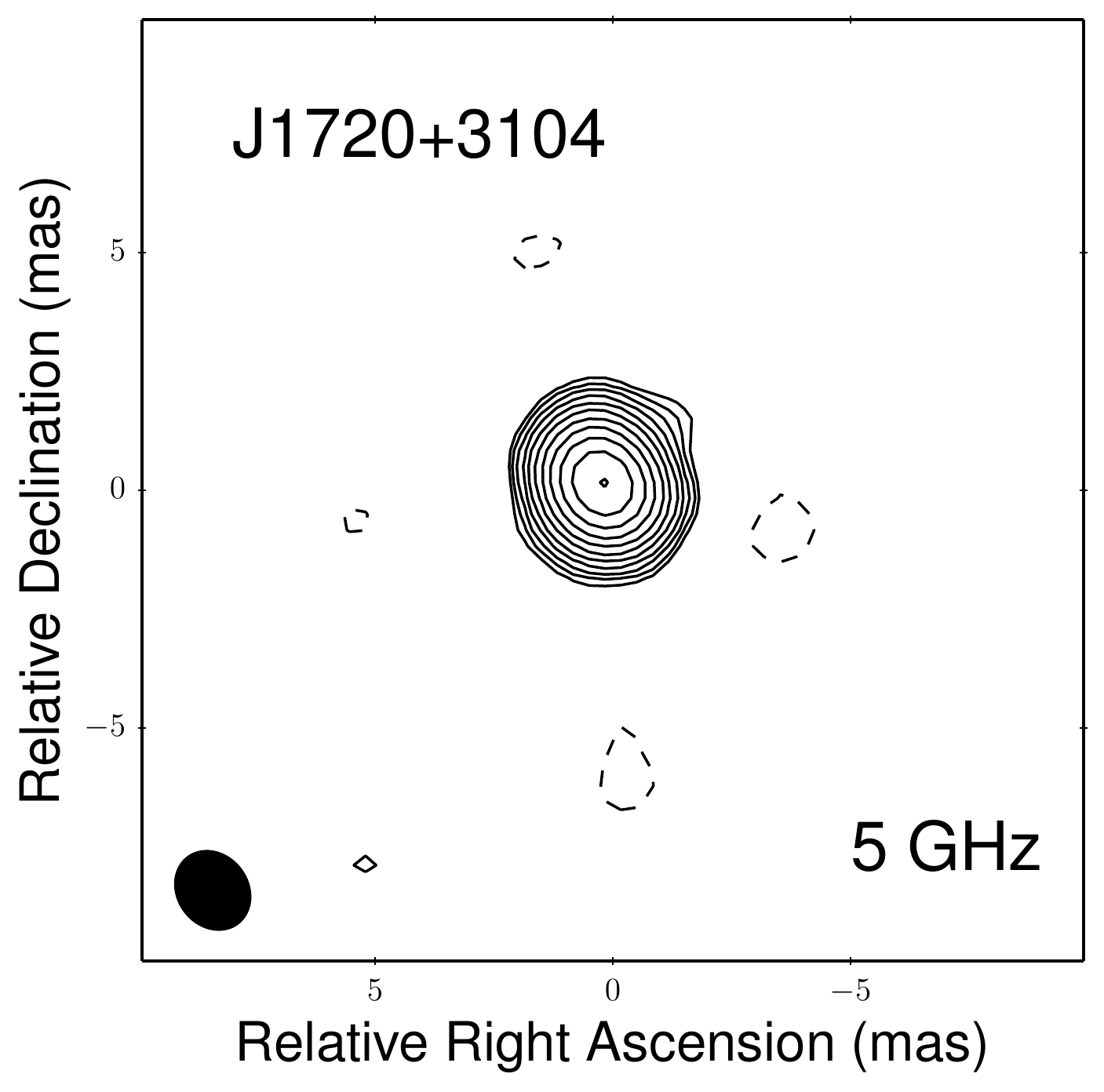}
  \end{minipage}}\\[-2ex]
  \vspace{0.1cm}
  \subfloat{
  \begin{minipage}{72.5mm}
    \centering
    \includegraphics[width=\columnwidth]{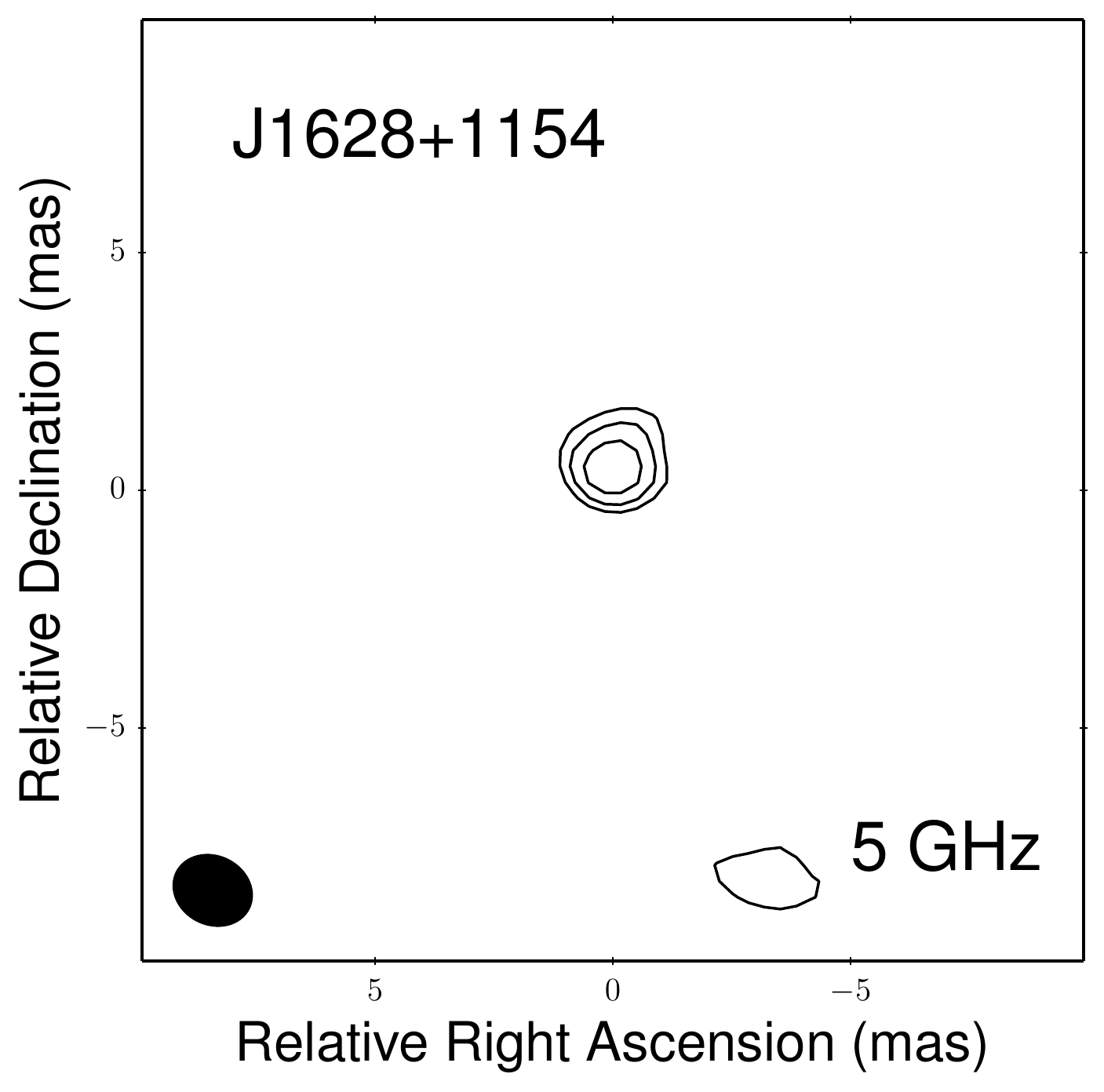}
  \end{minipage}  
  \begin{minipage}{72.5mm}
    \centering
    \includegraphics[width=\columnwidth]{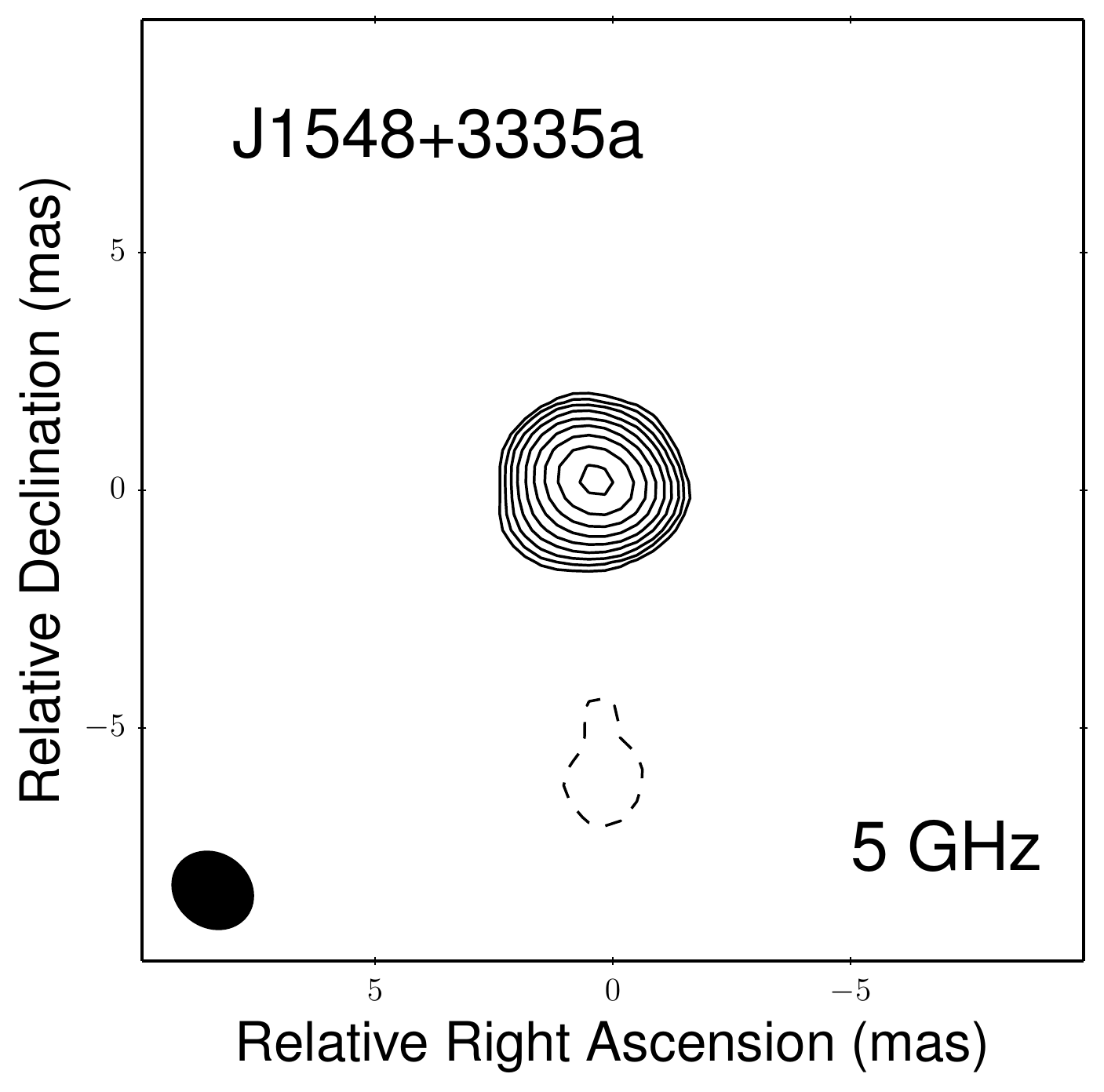}
  \end{minipage}}\\[-2ex]
  \contcaption{}
\end{figure*}

\begin{figure*}
  \subfloat{
  \begin{minipage}{75mm}
    \centering
    \includegraphics[width=\columnwidth]{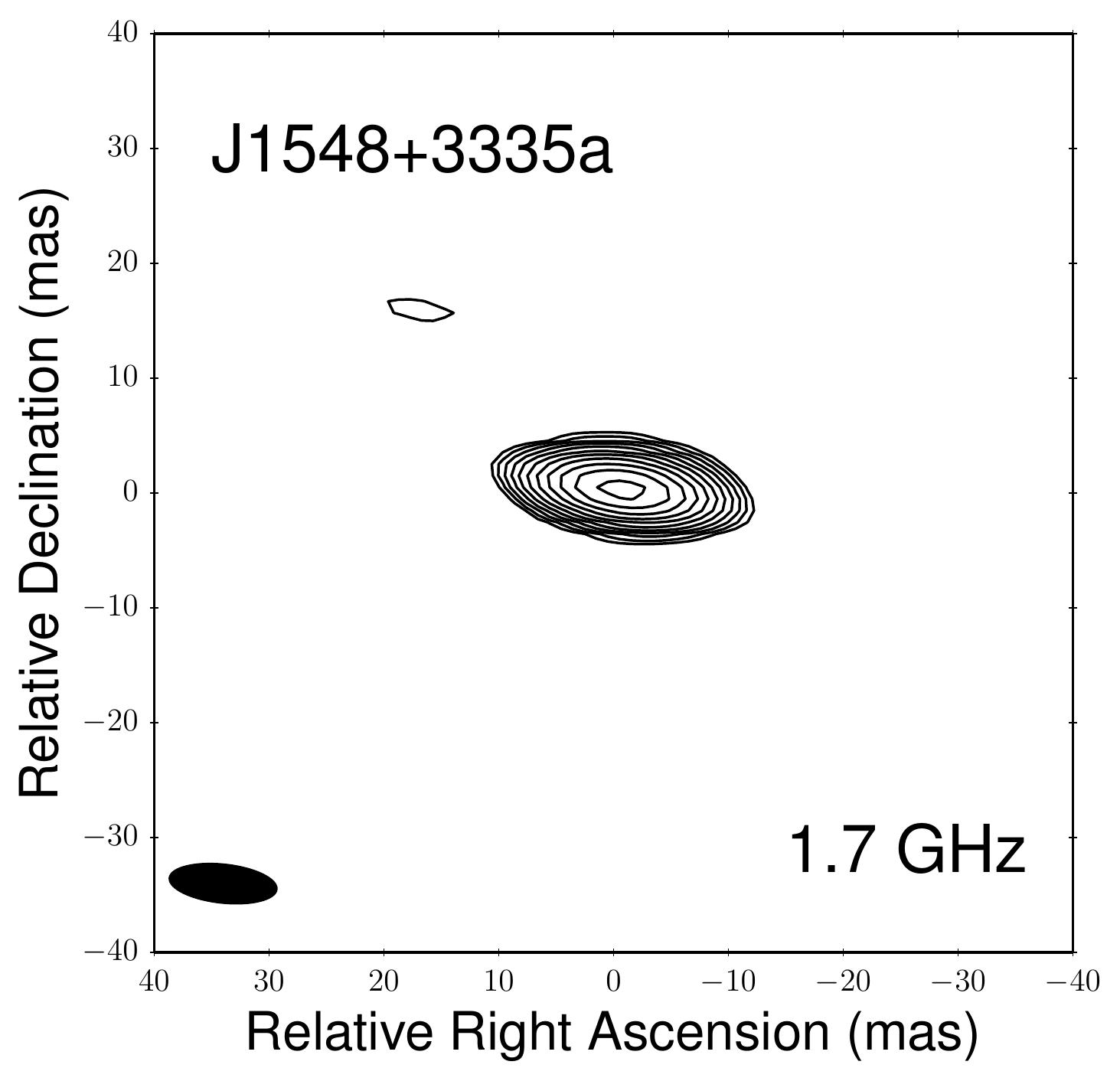}
  \end{minipage}  
  \begin{minipage}{73.5mm}
    \centering
    \includegraphics[width=\columnwidth]{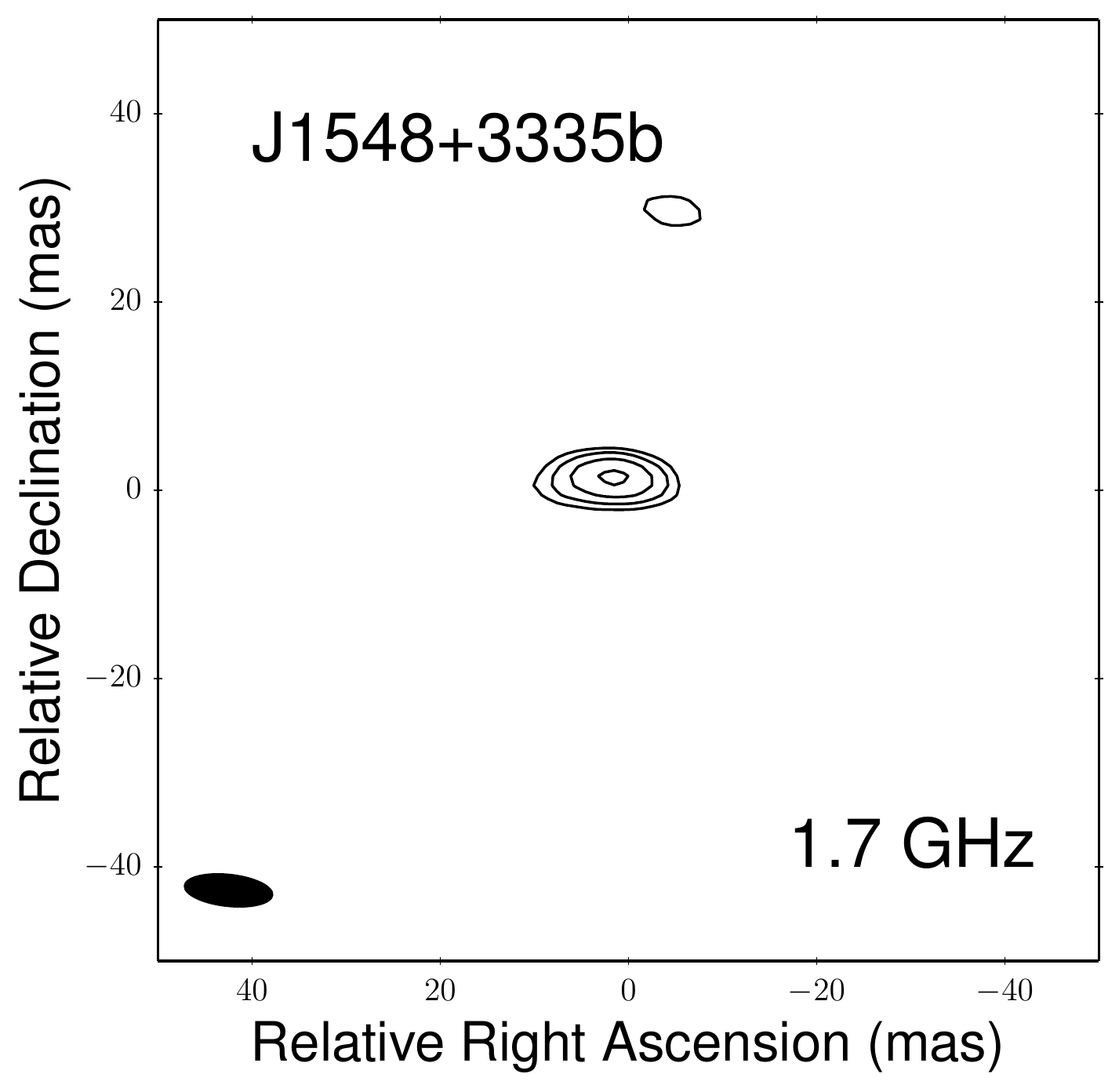}
  \end{minipage}}\\[-2ex]
  \vspace{0.1cm}
  \subfloat{
  \begin{minipage}{90mm}
    \hspace{-1cm}
    \includegraphics[width=\columnwidth]{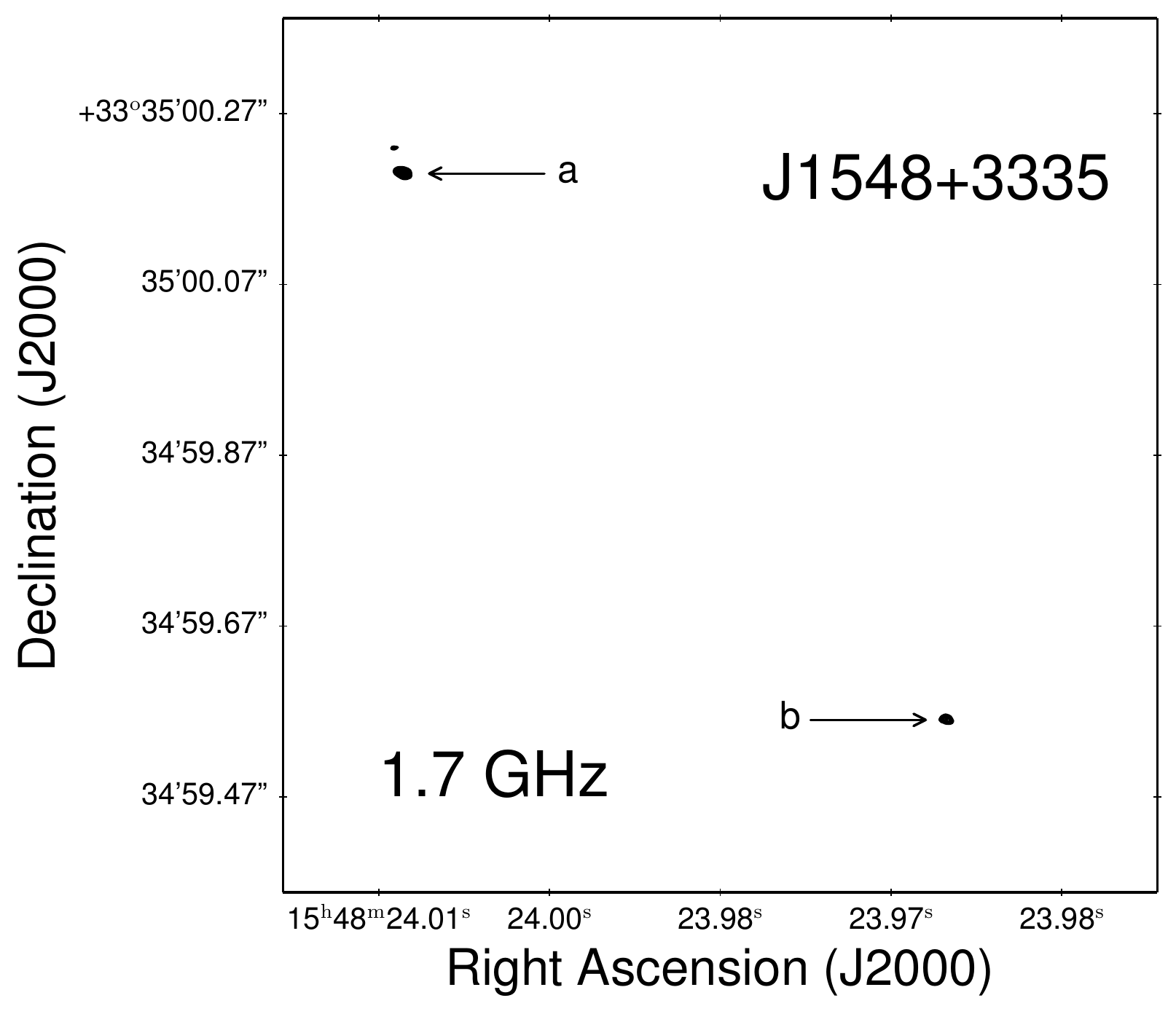}
  \end{minipage}  
  \begin{minipage}{0mm}
  \end{minipage}}\\[-2ex]
  \contcaption{}
\end{figure*}

\section{Results}
\label{sec:results}
The flux densities and sizes of the sources were calculated by fitting circular Gaussian brightness distribution models to the source visibility data in \textsc{difmap}. The parameters derived for each of the sources are presented in Table~\ref{tbl:source parameters}. The uncertainties were calculated using the equations in \citet{chap14whitebible}. An additional five\,per\,cent uncertainty of the flux densities were also assumed to account for the EVN amplitude calibration uncertainty \citep[e.g.][]{An2012,Frey2015}. The spectral index, $\alpha$, in Col.~9 is defined as $S \propto \nu^{\alpha}$, where $S$ is the flux density and $\nu$ is the frequency. The right ascension (RA) and declination (DEC) of the sources were calculated from the 5\,GHz images using the \textsc{aips} task \textsc{maxfit}. For J1548+3335b, which was not detected at 5\,GHz, the coordinates were derived from the 1.7\,GHz image. The uncertainties of RA and DEC were derived by adding the contributions from (1) the statistical uncertainty of the position of the source (which is a function of the angular resolution of the EVN in the given direction and the signal-to-noise ratio), (2) the uncertainty of the position of the phase-reference calibrator, and (3) the uncertainty introduced by the angular separation between the calibrator and the target source. 

All of the sources have fitted sizes that are smaller than the Gaussian restoring beam (Table \ref{tbl:image parameters}) except J1400+3149 at 5\,GHz and J1548+3335b at 1.7\,GHz. To check whether the remaining sources are resolved, we calculated the minimum resolvable size of each of the sources using Eq. 2 in \citet{2005AJ....130.2473K}. All of the sources and source components are resolved at both frequencies, except J0210$-$0018. This object remains unresolved with the array at 1.7\,GHz but it is resolved at 5\,GHz in our observations.

The redshift-corrected 5\,GHz brightness temperatures of the sources were calculated using
\begin{equation}
 T_{\mathrm b} = 1.22\times10^{12}(1+z)\frac{S}{\theta^2\nu^2}\,\,{\mathrm K}
 \label{eq:Tb}
\end{equation}
\citep{Condon1982}. Here, $z$ is the redshift, $S$ is the integrated flux density in Jy, $\theta$ is the fitted circular Gaussian full width at half-maximum (FWHM) diameter in mas, and $\nu$ is the observing frequency in GHz. 

\begin{table*}
 \centering
 \begin{minipage}{\textwidth}
  \caption{Source parameters}
  \begin{tabular}{ccccccccc}
  \hline
  ID   & $z$ & \multicolumn{5}{c}{1.7\,GHz flux density [mJy]} & 5\,GHz flux density & $\alpha$\,$^{\mathrm b}$\\
  \cline{3-7}
       &     & EC052A & EC052C & EC052E & EC052F & combined & [mJy] & \\
   (1) & (2) & (3)      & (4)      & (5)      & (6)      & (7)      & (8) & (9)\\
  \hline
  J0011+1446  & 4.96 & $19.3\pm1.0$                & $17.9\pm1.1$                & ---          & ---          & $18.6\pm1.0$                & $10.3\pm0.6$                & $-0.54\pm0.07$\\  
  J0210$-$0018  & 4.65 & $2.0\pm0.2$\,$^{\mathrm c}$ & $1.2\pm0.2$\,$^{\mathrm c}$ & ---          & ---          & $1.7\pm0.2$\,$^{\mathrm c}$ & $2.8\pm0.2$\,$^{\mathrm c}$ & $0.47\pm0.11$\\  
  J0940+0526  & 4.50 & $12.8\pm0.8$                & ---                         & $19.0\pm2.7$ & $21.5\pm1.1$ & $18.3\pm1.0$                & $14.0\pm0.7$                & $-0.24\pm0.07$\\  
  J1013+2811  & 4.75 & $9.4\pm0.5$                 & $11.6\pm0.7$                & ---          & $9.6\pm0.5$  & $10.4\pm0.5$                & $6.2\pm0.3$\,$^{\mathrm c}$ & $-0.47\pm0.07$\\  
  J1311+2227  & 4.61 & $4.1\pm0.2$\,$^{\mathrm c}$ & $3.4\pm0.2$\,$^{\mathrm c}$ & ---          & ---          & $3.6\pm0.2$\,$^{\mathrm c}$ & $1.7\pm0.1$\,$^{\mathrm c}$ & $-0.65\pm0.07$\\  
  J1400+3149  & 4.64 & ---                         & ---                         & $10.7\pm0.5$ & ---          &   ---                       & $5.6\pm0.7$\,$^{\mathrm c}$ & $-0.59\pm0.12$\\  
  J1454+1109  & 4.93 & ---                         & ---                         & $19.7\pm1.0$ & ---          &  ---                        & $28.6\pm1.4$                & $0.34\pm0.06$\\  
  J1548+3335a & 4.68 & $7.1\pm0.4$                 & ---                         & ---          & $7.4\pm0.4$  & $7.7\pm0.4$                 & $3.7\pm0.2$\,$^{\mathrm c}$ & $-0.66\pm0.07$\\  
  J1548+3335b & 4.68 & $0.6\pm0.1$                 & ---                         & ---          & $2.2\pm1.0$  & $2.3\pm1.6$                 & ---                         & ---\\  
  J1628+1154  & 4.47 & ---                         & ---                         & ---          & ---          & $<0.6^{\mathrm a,c}$          & $0.7\pm0.1$\,$^{\mathrm c}$ & $>0.15\pm0.26^{\mathrm a}$\\  
  J1720+3104  & 4.62 & $14.3\pm0.8$                & ---                         & $10.1\pm0.5$ & ---          & $10.9\pm0.6$                & $16.3\pm0.8$                & $0.36\pm0.07$\\  
  \hline
  \end{tabular}
  \label{tbl:source parameters}
 \end{minipage}
\end{table*}

\begin{table*}
 \hspace{-2cm}
 \centering
 \begin{minipage}{\textwidth}
  \contcaption{}
  \begin{tabular}{cccccccc}
  \hline
  ID   & RA & DEC & \multicolumn{2}{c}{Size [mas]} & \multicolumn{2}{c}{$T_{\mathrm b}$ [$10^9$\,K]} & $L$ \\
  \cline{4-5}
  \cline{6-7}
       & [J2000] & [J2000] & 1.7\,GHz & 5\,GHz & 1.7\,GHz & 5\,GHz & [$10^{26}$\,W\,Hz$^{-1}$] \\
   (1) & (10) & (11) & (12) & (13) & (14) & (15) & (16)\\
  \hline
  J0011+1446   & 00:11:15.23392 (0.07) & 14:46:01.8116 (1.0)  & $2.73\pm0.02$   & $0.82\pm0.01$ & $6.64\pm0.35$                       & $4.53\pm0.27$                  & $11.54\pm1.53$ \\  
  J0210$-$0018 & 02:10:43.16430 (0.07) & $-$00:18:18.4449 (1.6) & $<1.09\,^{\mathrm e}$       & $0.46\pm0.01$ & $>3.54\pm0.36$\,$^{\mathrm{c,d}}$ & $3.75\pm0.26$\,$^{\mathrm c}$  & $0.49\pm0.09$  \\  
  J0940+0526   & 09:40:04.80070 (0.08) & 05:26:30.9478 (1.9)  & $1.91\pm0.02$   & $0.63\pm0.01$ & $12.29\pm0.67$                      & $9.68\pm0.52$                  & $7.83\pm0.98$  \\  
  J1013+2811   & 10:13:35.44036 (0.07) & 28:11:19.2424 (1.0)  & $1.31\pm0.01$   & $0.59\pm0.01$ & $15.52\pm0.80$                      & $5.02\pm0.31$\,$^{\mathrm c}$  & $5.66\pm0.75$  \\  
  J1311+2227   & 13:11:21.32190 (0.07) & 22:27:38.6313 (1.0)  & $2.38\pm0.04$   & $1.08\pm0.03$ & $1.57\pm0.10$\,$^{\mathrm c}$       & $0.41\pm0.03$\,$^{\mathrm c}$  & $2.06\pm0.29$  \\  
  J1400+3149   & 14:00:25.41675 (0.07) & 31:49:10.6764 (1.0)  & $3.01\pm0.02$   & $2.69\pm0.28$ & $2.97\pm0.16$                       & $0.21\pm0.04$\,$^{\mathrm c}$  & $6.04\pm1.40$  \\  
  J1454+1109   & 14:54:59.30513 (0.08) & 11:09:27.8855 (1.4)  & $0.94\pm0.01$   & $0.46\pm0.01$ & $59.30\pm3.00$                      & $38.52\pm1.95$                 & $6.66\pm0.84$  \\  
  J1548+3335a  & 15:48:24.01400 (0.08) & 33:35:00.0862 (1.4)  & $1.76\pm0.02$   & $0.77\pm0.01$ & $6.28\pm0.33$                       & $1.76\pm0.11$\,$^{\mathrm c}$  & $4.60\pm0.60$  \\ 
  J1548+3335b  & 15:48:23.95861 (0.14) & 33:34:59.6640 (2.1)  & $20.58\pm14.03$ & ---           & ---                                 & ---                            & ---            \\  
  J1628+1154   & 16:28:30.46537 (0.07) & 11:54:03.4658 (1.1)  & ---             & $1.40\pm0.24$ & ---                                 & $0.09\pm0.03$\,$^{\mathrm c}$  & ---            \\  
  J1720+3104   & 17:20:26.68897 (0.07) & 31:04:31.6451 (1.2)  & $1.18\pm0.01$   & $0.51\pm0.01$ & $19.57\pm1.05$                      & $17.47\pm0.94$                 & $3.35\pm0.42$  \\  
  \hline
  \multicolumn{8}{p{18cm}}{\footnotesize{\textbf{Columns:} Col.~1 -- source name (J2000); Col.~2 -- redshift; Cols.~3 to 7 -- fitted 1.7\,GHz flux densities and uncertainties; Col.~8 -- fitted 5\,GHz flux density and uncertainty; Col.~9 -- 1.7 to 5\,GHz spectral index; Col.~10 -- right ascension and uncertainty in ms (in brackets); Col.~11 -- declination and uncertainty in mas (in brackets); Cols.~12 and 13 -- fitted 1.7 and 5\,GHz circular Gaussian diameter (FWHM); Col.~14 and 15 -- 1.7 and 5\,GHz brightness temperature; Col.~16 -- monochromatic rest-frame 5\,GHz luminosity. }}\\
  \multicolumn{8}{p{18cm}}{\footnotesize{\textbf{Notes:} $^{\mathrm a}$ J1628+1154 was not detected at 1.7\,GHz, the quoted limit corresponds to the $6\sigma$ detection threshold. $^{\mathrm b}$ For the sources that were observed more than once at 1.7\,GHz, the value was calculated using the combined 1.7\,GHz flux density. $^{\mathrm c}$ Phase self-calibration was not performed. See Section \ref{subsec:p-selfcal} for a discussion. $^{\mathrm d}$ The value is a lower limit since the source is unresolved. $^{\mathrm e}$ Since the source is unresolved, the value is the minimum resolvable size calculated using Eq. 2 in \citet{2005AJ....130.2473K}.}}\\
  \end{tabular}
  \label{tbl:source parameters cont}
 \end{minipage}
\end{table*}

\subsection {Phase self-calibration}
\label{subsec:p-selfcal}
In Section~\ref{sec:obsrve and reduce} we noted that phase-only self-calibration was only applied to the sources in which the sum of the \textsc{clean} component flux densities exceeded $\sim10$\,mJy. The sources for which phase-only self-calibration was not applied are marked in Table~\ref{tbl:source parameters}. The flux densities of these sources may be underestimated by about 10\,per\,cent \citep[e.g.][]{2010A&A...515A..53M,2010A&A...524A..83F}. Consequently, the brightness temperatures and luminosities of these sources could also be about 10\,per\,cent higher than indicated in Table~\ref{tbl:source parameters}.

\subsection{1.7\,GHz flux densities}
\label{subsec:multi epoch}
In this section we check the consistency of the flux densities of the sources that were observed in more than one different project segments at 1.7\,GHz. Among these sources, all the flux densities of {\bf J0011+1446} and {\bf J1548+3335a} are consistent with each other within their formal errors (Table~\ref{tbl:source parameters}). This is not the case for {\bf J0210$-$0018}, {\bf J0940+0526}, {\bf J1013+2811}, {\bf J1311+2227}, {\bf J1548+3335b}, and {\bf J1720+3104}. Therefore we cannot exclude that these sources are variable. Another possibility is that the sources are not variable but the errors of the fitted flux densities are somewhat underestimated. The flux density errors are dominated by the assumed 5\,per\,cent calibration uncertainty. Since it is not possible to do absolute flux density calibration during VLBI experiments using standard calibrator sources, it may be that the systematic uncertainty in the flux densities is occasionally larger than five\,per\,cent. Increasing the assumed systematic uncertainty to 10\,per\,cent, all of the 1.7\,GHz flux densities of J1013+2811 and J1311+2227 also become consistent with each other within their errors. In addition, the quoted errors are $1\sigma$ uncertainties. This means that there is a 32\,per\,cent chance that the flux density lies outside of the indicated error bar. Taking all of the above into account, we consider it unlikely that J0210$-$0018 is variable since the flux densities between the observations differ by at most 0.8\,mJy. For J1548+3335b, the EC052F and the combined flux densities are consistent within their uncertainties and inconsistent with the EC052A flux density. Considering how faint and extended the component is, it is unlikely that it is variable. The flux density of J0940+0526 measured in the project segment EC052A is lower than the values from EC052E, EC052F and the combined data set. For J1720+3104, the EC052A flux density is higher than the EC052F and the combined values. It is therefore possible that J0940+0526 and J1720+3104 are indeed variable.

Since the combined data sets that include data from all project segments have better $(u,v)$ coverage and sensitivity than the individual data sets, these were used for imaging (Fig.~\ref{fig:VLBI images}), and the flux densities from the combined data sets were used to calculate the relevant parameters in Table~\ref{tbl:source parameters}. Unless specified otherwise, the discussions in this paper will be based on the results from the combined 1.7\,GHz datasets. 

\subsection{Comments on peculiar sources}
\label{subsec:Comments on Individual Sources}
{\bf J1548+3335.} The 1.7\,GHz VLBI image of the source shows two components separated by $812\pm3$\,mas, corresponding to a projected linear separation of $5267\pm17$\,pc (Fig.~\ref{fig:VLBI images}). This source structure is reminiscent of that of the medium-size symmetric objects (MSOs) \citep[e.g.][]{Fanti1995}. The fainter component, J1548+3335b, is not detected in the naturally weighted 5\,GHz image with a local 6$\sigma$ noise of $0.18$mJy\,beam$^{-1}$. Considering how faint J1548+3335b is and that it has a size of $20.58\pm14.03$\,mas, we conclude that it was resolved out by the higher resolution 5\,GHz observations. It is also likely that its spectrum is steep therefore its flux density is lower at 5\,GHz than at 1.7\,GHz. J1548+3335a is an AGN (see Section \ref{subsec:agn star}). Because of the large uncertainty on the size of J1548+3335b, the emission from this component could either be from an AGN or star formation (Section \ref{subsec:agn star}). The SDSS source is $\sim60$\,mas away from J1548+3335a, indicating that this radio component positionally coincides with the optical AGN, since SDSS sources have a positional uncertainty of $\sim60$\,mas \citep[e.g.][]{2013A&A...553A..13O}. 

There are in principle four possibilities for what J1548+3335b could be: (1) it is an unrelated foreground or background AGN or star-forming source; (2) J1548+3335b is a lobe or hotspot in the kpc-scale extended radio structure of J1548+3335a; (3) J1548+3335 is a kpc-scale separation dual AGN system; (4) J1548+3335a and J1548+3335b are gravitationally lensed images of the same source. If J1548+3335a and J1548+3335b are produced by gravitational lensing they should have the same spectral index, in which case J1548+3335b is predicted to have a 5\,GHz flux density of $1.1\pm0.8$\,mJy. Consequently, if J1548+3335b is compact, it should have been detected in our experiment at 5\,GHz. However, because of the large uncertainty on its size, it is still possible that it was resolved out by the higher resolution 5\,GHz observations. Therefore we cannot rule out the possibility of gravitational lensing, although the positional coincidence of the corresponding SDSS quasar with J1548+3335a is at odds with the scenario where the optical object should be the blend of the two putative lensed images. A straightforward way to test if J1548+3335a and J1548+3335b are related would be to do lower resolution, higher sensitivity radio interferometric observations of the system. Such observations could detect the emission from a jet or a flat spectrum core between the components. The former confirming that J1548+3335b is a lobe or hotspot of J1548+3335a and the latter that J1548+3335 is a MSO, ruling out the other possibilities. If no emission is detected between the components, this could also allow a spectral index to be calculated for J1548+3335b, thereby refuting or strengthening the possibility that J1548+3335a and J1548+3335b are gravitationally lensed images of the same source.

Calculating the spectral index of J1548+3335 using the sum of the flux densities of the 1.7\,GHz components (rather than calculating the spectral indices separately for each component, as was done in Table~\ref{tbl:source parameters}) gives $\alpha_{\mathrm{source}}  = -1.07\pm0.07$. By extrapolating the power-law spectrum using $\alpha_{\mathrm{source}}$ we calculated a predicted 1.4\,GHz flux density ($S_{\mathrm{pred}}$) from our high-resolution VLBI data. From that, we calculated a 1.4\,GHz flux density ratio with FIRST, $S_{\mathrm{pred}}/S_{\mathrm{FIRST}} = 0.38\pm0.06$. Assuming that J1548+3335 is not variable, this indicates that most of its radio emission is resolved out by the EVN and strengthens the possibility that some of the missing flux density is located in a jet between the two components or in other extended structures such as radio lobes.

As a result of the large uncertainty of the size of J1548+3335b and the brightness temperature being inversely proportional to the size of the source squared, the formal uncertainty of the J1548+3335b brightness temperature becomes twice as large as the value itself. Hence, no value is shown in Table~\ref{tbl:source parameters}.

\noindent
{\bf J1628+1154.} This source is not detected in the naturally weighted 1.7\,GHz image with 0.6\,mJy\,beam$^{-1}$ (6$\sigma$) noise. J1628+1154 is detected in the higher resolution 5\,GHz image with a flux density slightly above the detection threshold of the 1.7\,GHz image. This suggests a positive spectral index or perhaps flux density variability over the time scale of several months.

\section{VLBI images of z$>$4.5 sources from the literature}
\label{sec:vlbi_z>4.5}
To the best of our knowledge, twenty $z>4.5$ sources have been imaged with VLBI prior to this paper. These were collected from the literature and the Optical Characteristics of Astrometric Radio Sources (OCARS) catalogue\footnote{http://www.gao.spb.ru/english/as/ac\_vlbi/ocars.txt}\citep{2008mefu.conf..183M,2009A&A...506.1477T}. Table~\ref{tbl:vlbi_z>4.5} contains a summary of these sources. If a source is composed of two or more components, the integrated flux density in Col.~4 is the sum of the flux densities of the components. The spectral indices in Col.~5 were calculated using the flux densities in Col.~4, and are therefore the spectral indices of the sources as a whole rather than individual component spectral indices (as is the case in Table~\ref{tbl:source parameters}). Column~6 contains a visual classification of each of the sources. A source is marked with S if its VLBI image shows a single component that is symmetric with no appreciable extension. Extended (E) sources have single components that are asymmetric with respect to one of their axes, and are therefore resolved. Multi-component (MC) sources have two or more distinct components. The brightness temperatures (Col.~7) and monochromatic rest-frame luminosities (Col.~8) are either from the VLBI reference given in Col.~9 or are calculated in the same way as described in Section~\ref{sec:results}. In case of multi-component sources, both the brightness temperatures and luminosities are of the main (brightest) source component. For these sources, the luminosity is calculated using the spectral index of the main source component, where the spectral index is derived between the same frequencies as the source spectral index reported in Col.~5. Brightness temperatures of unresolved sources and source components are indicated as lower limits.

\begin{table*}
 \hspace*{-1.5cm}
 \centering
 \begin{minipage}{\textwidth}
  \caption{Sources at $z>4.5$ that have previously been imaged with VLBI.}
  \begin{tabular}{ccccccccc}
  \hline
  ID           & $z$& $\nu$ & Flux density$^{\mathrm f}$ & $\alpha_{\mathrm{source}}$\,$^{\mathrm b}$ & Classification$^{\mathrm d}$ & $T_{\mathrm b}$ & $L$                       & VLBI \\
               &    & [GHz] & [mJy]                       &                                           &                               &         [K]     & [$10^{26}$\,W\,Hz$^{-1}$] & ref.\\
  (1)          & (2)& (3)   & (4)                         & (5)                                       & (6)                           & (7)             & (8)                       & (9) \\
  \hline  
  J0131$-$0321 & 5.18 & 1.7 & $64.4\pm0.3$        & ---                       & S  & $>2.8\times10^{11}$\,$^{\mathrm e}$           & $>30.00$\,$^{\mathrm e}$ & 11\\ 
  \hline
  J0311+0507   & 4.51 & 1.7 & 417.7               & $-1.48^{(1.7,5)}$         & MC & $>2.0\times10^7$\,$^{\mathrm{e,g}}$           & $>11.88\,^{\mathrm e,g}$ & 19\\ 
               &      & 5   & 82.0                &                           & MC & ---                                          & --- & 19\\
  \hline
  J0324$-$2918 & 4.63 & 2.3 & $214.0\pm10.9$ & $-0.48\pm0.06^{(2.3,8.7)}$ & S & $(1.13\pm0.06)\times10^{11}$ & $188.70\pm21.68$ & 22 \\
               &      & 8.6 & $113.8\pm6.9$  &                            & S & $(6.8\pm0.5)\times10^{10}$   & $100.34\pm11.98$ & 22 \\
  \hline
  J0813+3508   & 4.92 & 1.6 & $17.1\pm0.9$        & $-0.77\pm0.07^{(1.6,5)}$  & MC & ---                                          & $15.04\pm1.71$ & 10\\
               &      & 5   & $7.3\pm0.4$         &                           & E  & $(1.5\pm0.1)\times10^9$                      & $8.40\pm0.40$ & 10\\
  \hline
  J0836+0054   & 5.77 & 1.6 & 1.1                 & $-1.03^{(1.6,5)}$         & S  & $>3.2\times10^7$\,$^{\mathrm e}$             & $>4.25$\,$^{\mathrm e}$ & 6\\ 
               &      & 5   & 0.34                &                           & S  & $>3.4\times10^6$\,$^{\mathrm e}$             & $>1.40$\,$^{\mathrm e}$ & 7\\ 
  \hline
  J0906+6930   & 5.47 & 15  & $121.3\pm0.5$       & $-0.94\pm0.05^{(15,43)}$  & MC & $>5.3\times10^9$\,$^{\mathrm e}$             & $>351.77\pm28.79$\,$^{\mathrm e}$ & 5\\ 
               &      & 43  & $46.1\pm2.2$        &                           & MC & $>1.1\times10^9$\,$^{\mathrm e}$             & $>128.47\pm12.01$\,$^{\mathrm e}$ & 5\\ 
  \hline
  J0913+5919   & 5.11 & 1.4 & $19.4\pm0.1$        & ---                       & S  & $>4.2\times10^{10}$\,$^{\mathrm e}$           & --- & 1\\ 
  \hline
  J1026+2542   & 5.27 & 1.7 & $180.4\pm5.0$       & $-0.75\pm0.04^{(1.7,5)}$  & E  & $>(2.3\pm0.1)\times10^{12}$\,$^{\mathrm e}$  & $>56.01\pm7.26$\,$^{\mathrm e}$ & 18\\  
               &      & 4.9 & $69.6\pm2.5$        &                           & MC & $(1.7\pm0.1)\times10^{10}$                   & --- & 12\,\&\,13\\
               &      & 5   & $79.2\pm2.9$        &                           & MC & $>(5.7\pm0.3)\times10^{11}$\,$^{\mathrm e}$  & $>28.65\pm3.70$\,$^{\mathrm e}$ & 18\\  
  \hline
  J1146+4037   & 5.01 & 1.6 & $15.5\pm0.8$        & $-0.53\pm0.06^{(1.6,5)}$  & S  & ---                                          & $18.28\pm2.29$ & 10\\
               &      & 5   & $8.6\pm0.4$         &                           & E  & $(4.5\pm0.3)\times10^9$                      & $10.70\pm0.50$ & 10\\
  \hline
  J1205$-$0742 & 4.69 & 1.4 & $0.57\pm0.05$       & ---                       & MC & $(2.2\pm0.3)\times10^4$                      & --- & 2\\
  \hline
  J1235$-$0003 & 4.69 & 1.4 & $17.2\pm0.1$        & ---                       & S  & $>5.4\times10^9$\,$^{\mathrm e}$             & --- & 1\\ 
  \hline
  J1242+5422   & 4.73 & 1.6 & $17.7\pm0.9$        & $-0.55\pm0.07^{(1.6,5)}$  & E  & ---                                          & $17.59\pm2.28$ & 10\\
               &      & 5   & $9.7\pm0.5$         &                           & S  & $(5.9\pm0.5)\times10^9$                      & $9.00\pm0.50$ & 10\\
  \hline
  J1427+3312   & 6.12 & 1.4 & $1.8\pm0.1$         & $-1.06^{(1.6,5)}$         & MC & $3.9\times10^8$                              & --- & 8\\ 
               &      & 1.6 & 1.5                 &                           & MC & $9.4\times10^7$                              & 1.78 & 9\\
               &      & 5   & 0.46                &                           & E  & $1.5\times10^7$                              & 0.89 & 9\\
  \hline
  J1429+5447   & 6.21 & 1.6 & $3.3\pm0.1$         & $-1.09\pm0.06^{(1.7,5)}$  & S  & $(1.40\pm0.06)\times10^9$                    & $13.48\pm1.53$ & 16\\
               &      & 5   & $0.99\pm0.06$       &                           & E  & $(7.7\pm0.7)\times10^8$                      & $4.50$ & 16\\
  \hline
  J1430+4204   & 4.72 & 2.3 & 232.0$^{\mathrm a}$ & $0.17\&-0.04^{(5,15),\mathrm c}$ & E & ---                                    & --- & 14\,\&\,15\\
               &      & 5   & 173.0               &                           & S  & $2.7\times10^{11}$                           & $50.92\, \& \,73.40$\,$^{\mathrm c}$ & 3\\
               &      & 8.6 & 111.0$^{\mathrm a}$ &                           & E  & ---                                          & --- & 14\,\&\,15\\
               &      & 15  & $209.0\&166.0^{\mathrm c}$ &                    & E  & $4.3\&6.0\times10^{11}$\,$^{\mathrm c}$      & $61.52\,\&\,70.43$\,$^{\mathrm c}$ & 4\\  
  \hline
  \end{tabular}
  \label{tbl:vlbi_z>4.5}
 \end{minipage}
\end{table*}

\begin{table*}
 \hspace*{-0.9cm}
 \centering
 \begin{minipage}{\textwidth}
  \contcaption{}
  \begin{tabular}{ccccccccc}
  \hline
  ID           & $z$& $\nu$ & Flux density$^{\mathrm f}$ & $\alpha_{\mathrm{source}}$\,$^{\mathrm b}$ & Classification$^{\mathrm d}$ & $T_{\mathrm b}$ & $L$                       & VLBI \\
               &    & [GHz] & [mJy]                       &                                           &                               &         [K]     & [$10^{26}$\,W\,Hz$^{-1}$] & ref.\\
  (1)          & (2)& (3)   & (4)                         & (5)                                       & (6)                           & (7)             & (8)                       & (9) \\
  \hline
  J1606+3124   & 4.56 & 2.3 & $952.1\pm48.2$      & $-0.38\pm0.09^{(2.3,4.8)}$& S  & $(3.8\pm0.2)\times10^{11}$                   & $908.45\pm154.23$ & 20\\
               &      & 4.8 & $712.0\pm32.8$      &                           & MC & $(2.2\pm0.1)\times10^{11}$                   & $602.30\pm102.27$ & 13\\
               &      & 8.3 & $471.7\pm23.4$      &                           & MC & $(5.8\pm0.3)\times10^{10}$                   & --- & 20\\
  \hline
  J1611+0844   & 4.54 & 1.6 & $13.0\pm0.8$        & $-0.01\pm0.07^{(1.6,5)}$  & S  & ---                                          & $5.24\pm0.71$ & 10\\
               &      & 5   & $12.9\pm0.6$        &                           & E  & $(4.7\pm0.3)\times10^9$                      & $4.60\pm0.20$ & 10\\
  \hline
  J1659+2101   & 4.78 & 1.6 & $29.3\pm1.5$        & $-0.92\pm0.08^{(1.6,5)}$  & E  & ---                                          & $21.41\pm3.04$ & 10\\
               &      & 5   & $10.6\pm0.7$        &                           & E  & $(3.0\pm0.4)\times10^8$                      & $12.30\pm0.80$ & 10\\
  \hline
  J2102+6015   & 4.58 & 2.3 & $298.6\pm14.9$ & $-0.99\pm0.07^{(2.3,8.3),\,\mathrm{j}}$ & S & $(4.1\pm0.2)\times10^{10}$  & $618.26\pm77.67$\,$^{\mathrm h}$ & 20\\
               &      & 2.3 & $296.4\pm14.9$ & $-0.57\pm0.05^{(2.3,8.6),\,\mathrm{k}}$ & S & $(2.7\pm0.1)\times10^{10}$   & $296.42\pm31.45$\,$^{\mathrm i}$ & 21\\
               &      & 8.3 & $82.4\pm5.9$   &                            & S & $>(1.14\pm0.08)\times10^{11}$\,$^{\mathrm e}$   & $170.66\pm23.18$\,$^{\mathrm h}$ & 20\\
               &      & 8.6 & $140.6\pm7.2$  &                            & S & $(1.49\pm0.08)\times10^{10}$ & $140.57\pm15.00$\,$^{\mathrm i}$ & 21\\
  \hline
  J2228+0110   & 5.95 & 1.7 & $0.30\pm0.12$       & ---                       & S  & $>3.1\times10^8$\,$^{\mathrm e}$             & --- & 17\\ 
  \hline
  \multicolumn{9}{p{18cm}}{\footnotesize{\textbf{Columns:} Col.~1 -- source name (J2000); Col.~2 -- redshift; Col.~3 -- VLBI observing frequency; Col.~4 -- integrated VLBI flux density; Col.~5 -- source spectral index; Col.~6 -- visual classification of the compact radio structure; Col.~7 -- brightness temperature of the main (brightest) source component; Col.~8 -- monochromatic rest-frame luminosity at the VLBI observing frequency of the main (brightest) source component; Col.~9 -- VLBI literature reference.}}\\
  \multicolumn{9}{p{18cm}}{\footnotesize{\textbf{Notes:} $^{\mathrm a}$ The value is the peak brightness in mJy\,beam$^{-1}$. $^{\mathrm b}$ The frequencies (in GHz) between which the spectral indices were calculated are given as superscripts to the values. $^{\mathrm c}$ \citet{2010A&A...521A...6V} imaged the source twice to search for variability. The first value corresponds to the first image and the second value to the second image. $^{\mathrm d}$ S: single-component; E: extended; MC: multi-component. $^{\mathrm e}$ The value is a lower limit since the source or component (in the case of a multi-component source) is unresolved. $^{\mathrm f}$ For sources composed of multiple components, the quoted value is the sum of the flux densities of the components. $^{\mathrm g}$ \citet{2014MNRAS.439.2314P} found that J0311+0507 is composed of eight components. The flux density of the third component was used to calculate the value as the authors conclude that it is the core. $^{\mathrm h}$ The value was calculated using the spectral index between 2.3 and 8.3\,GHz. $^{\mathrm i}$ The value was calculated using the spectral index between 2.3 and 8.6\,GHz. $^{\mathrm j}$ The value was calculated using the \citet{2002ApJS..141...13B} 2.3\,GHz flux density. $^{\mathrm k}$ The value was calculated using the \citet{2008AJ....136..580P} 2.3\,GHz flux density.}}\\
  \multicolumn{9}{p{18cm}}{\footnotesize{\textbf{References:} 1: \citet{2004AJ....127..587M}; 2: \citet{2005AJ....129.1809M}; 3: \citet{1999A&A...344...51P}; 4: \citet{2010A&A...521A...6V} 5: \citet{2004ApJ...610L...9R}; 6: \citet{2003MNRAS.343L..20F}; 7: \citet{2005A&A...436L..13F} 8: \citet{2008AJ....136..344M} 9: \citet{2008A&A...484L..39F}; 10: \citet{2010A&A...524A..83F}; 11: \citet{2015MNRAS.450L..57G}; 12: \citet{2013MNRAS.431.1314F}; 13: \citet{2007ApJ...658..203H}; 14: \citet{2012ApJ...756L..20C}; 15: \citet{2004AJ....127.3587F}; 16: \citet{2011A&A...531L...5F}; 17: \citet{2014A&A...563A.111C}; 18: \citet{Frey2015}; 19: \citet{2014MNRAS.439.2314P}; 20: \citet{2002ApJS..141...13B}; 21: \citet{2008AJ....136..580P}; 22: \citet{2006AJ....131.1872P} }}\\
  \end{tabular}
  \label{tbl:vlbi_z>4.5 continued}
 \end{minipage}
\end{table*}

\subsection{Comments on peculiar literature sources}
\label{subsec:litrature Comments on Individual Sources}
{\bf J0311+0507.} The VLBI image of the source \citep[shown in ][]{2014MNRAS.439.2314P} has a complex structure with eight components. J0311+0507 has an angular size of 2.8\,arcsec which translates to a linear size of 18.7\,kpc.

\noindent
{\bf J0324$-$2918, J1606+3124 } and \textbf{ J2102+6015.} The sources are bright VLBI calibrators \citep{2002ApJS..141...13B,2006AJ....131.1872P,2008AJ....136..580P}. Since we could not find published brightness temperature or luminosity values, we downloaded the calibrated Very Long Baseline Array (VLBA) visibility data from the Astrogeo data base. The data were imaged and model-fitted in \textsc{difmap}, and the values in Table~\ref{tbl:vlbi_z>4.5} were calculated in the same way as described in Sections~\ref{sec:obsrve and reduce} and \ref{sec:results}.

\noindent
{\bf J1205$-$0742.} Based on its brightness temperature, the spectral index between 1.4 and 350\,GHz, linear size and morphology, \citet{2005AJ....129.1809M} concluded that J1205$-$0742 is a nuclear starburst without a radio-loud AGN.

\noindent
{\bf J1430+4204.} \citet{2010A&A...521A...6V} reported on VLBA observations of J1430+4204 made at two different epochs at 15\,GHz and found the source to be variable (Table~\ref{tbl:vlbi_z>4.5}), which is consistent with the results of the 15\,GHz total flux density monitoring \citep{1999MNRAS.308L...6F}.

\section{Discussion}
\label{sec:discus}
In this section, we discuss the origin of the radio emission, variability properties, spectral indices and Doppler boosting of the $z>4.5$ VLBI sources. The sample consists of the objects for which VLBI images are reported in this paper for the first time, and those found in the literature. A summary of the properties of each source is given in Table~\ref{tbl:Source clasification}.

\subsection{The origin of the radio emission}
\label{subsec:agn star}
Thermal emission from star formation typically has $T_{\mathrm{b}}<10^5$\,K \citep{Sramek1986,Condon1991,Kewley2000} while $T_{\mathrm{b}} \geq 10^6$\,K indicates non-thermal emission from AGN \citep[e.g.][]{Kewley2000,Middelberg2011}. Since all of the sources have $T_{\mathrm{b}} > 10^6$\,K, the expectation is that they are all AGN except {\bf J1205$-$0742} in which the radio emission is from a nuclear starburst (Section~\ref{subsec:litrature Comments on Individual Sources}). Non-thermal emission could, however, also originate from a supernova remnant or a nuclear supernova remnant complex \citep[e.g.][]{2012MNRAS.423.1325A}. Hence, brightness temperature on its own is not sufficient to prove that the emission is from an AGN.

Using optical and near-infrared spectral energy distributions of 942 1.4\,GHz radio sources, \citet{McAlpine2013} calculated luminosity functions and redshifts for star forming and AGN dominated radio galaxies. From this, \citet{Magliocchetti2014} showed that at $z>1.8$ sources with 1.4\,GHz radio luminosities smaller than $4\times10^{24}\mathrm{W\,Hz^{-1}}$ are powered by star formation while sources with radio luminosities larger than $4\times10^{24}\mathrm{W\,Hz^{-1}}$ are AGN powered. While it is possible that a source powered by star formation could be classified as AGN powered or vice versa, \citet{Magliocchetti2014} noted that the probability of this happening is very small. 

Using the spectral indices of the main source components we calculated predicted 1.4\,GHz flux densities for the main components of the $z>4.5$ sources and from that derived the 1.4\,GHz monochromatic rest-frame luminosities of the components. Using the smallest values of the luminosities allowed inside their uncertainties, the sources indicated in Table~\ref{tbl:Source clasification} are all powered by AGN. 1.4\,GHz luminosities could not be calculated for the remaining five sources because, for these sources, the flux densities of the main components are not available at both frequencies over which the spectral indices are calculated. It is therefore possible that the radio emission in these sources are form star formation. We do, however, consider it unlikely because of their large brightness temperatures, all of which are lower limits except in the case of {\bf J1628+1154}. This conclusion is supported by the result that the majority of extragalactic sources with 1.4\,GHz flux densities greater than $\sim10$\,mJy are AGN while star forming sources start to dominate at flux densities below $\sim1$\,mJy \citep[e.g.][and references therein]{1987ApJ...322L...9T,2010A&ARv..18....1D,2012ApJ...758...23C}.  

\subsection{Variability}
\label{subsec:variability}
Based on the available radio flux density data, we make an attempt to assess the variability of the sources. To check source variability in a uniform way, we matched all of the $z>4.5$ sources to the FIRST and the National Radio Astronomy Observatory (NRAO) VLA Sky Survey \cite[NVSS;][]{nvss} catalogues, and calculated the integrated flux density ratio ($S_{\mathrm{FIRST}}/S_{\mathrm{NVSS}}$) between the catalogue values (Table~\ref{tbl:Flux density ratios}). Both surveys were performed at the same observing frequency of 1.4\,GHz, using different configurations of the VLA. The resulting angular resolutions are $\sim5$\,arcsec (B configuration) and $\sim45$\,arcsec (D configuration) in FIRST and NVSS, respectively.  

\begin{table}
 \centering
 \begin{minipage}{\columnwidth}
  \caption{Flux density ratios of the $z>4.5$ sources}
  \begin{tabular}{ccc}
  \hline
  ID & $S_{\mathrm{FIRST}}/S_{\mathrm{NVSS}}$ & $S_{\mathrm{pred}}/S_{\mathrm{FIRST}}$ \\
  \hline  
  J0011+1446   & $0.68\pm0.04$ & $0.84\pm0.07$\\
  J0131$-$0321 & $1.07\pm0.06$ & ---\\
  J0210$-$0018 & $0.85\pm0.06$ & $0.18\pm0.02$\\
  J0311+0507   & $1.02\pm0.06$ & 1.05\\
  J0324$-$2918 & --- & $1.15\pm0.09$\,$^{\mathrm c}$\\ 
  J0813+3508   & $1.05\pm0.06$ & $0.52\pm0.04$\\
  J0836+0054   & $0.44\pm0.09$ & 1.14\\
  J0906+6930   & --- & ---\\ 
  J0913+5919   & $0.99\pm0.06$ & $1.11\pm0.06$\,$^{\mathrm b}$\\ 
  J0940+0526   & $0.95\pm0.06$ & $0.33\pm0.03$\\
  J1013+2811   & $0.91\pm0.06$ & $0.78\pm0.06$\\
  J1026+2542   & $0.93\pm0.05$ & $0.85\pm0.05$\\
  J1146+4037   & $1.00\pm0.06$ & $1.35\pm0.11$\\
  J1205$-$0742 & --- & $1.05\pm0.32$\,$^{\mathrm a}$\\
  J1235$-$0003 & $1.01\pm0.06$ & $0.93\pm0.05$\,$^{\mathrm b}$\\ 
  J1242+5422   & $1.03\pm0.06$ & $0.96\pm0.08$\\
  J1311+2227   & $0.87\pm0.07$ & $0.61\pm0.05$\\
  J1400+3149   & $0.94\pm0.06$ & $0.58\pm0.06$\\
  J1427+3312   & --- & 1.72\\
  J1429+5447   & $0.78\pm0.11$ & $1.35\pm0.09$\\
  J1430+4204   & $1.02\pm0.06$ & ---\\
  J1454+1109   & $1.54\pm0.11$ & $1.24\pm0.10$\\
  J1548+3335   & $0.99\pm0.06$ & $0.38\pm0.06$\\
  J1606+3124   & $1.03\pm0.06$ & $1.67\pm0.18$\\
  J1611+0844   & $1.01\pm0.08$ & $1.48\pm0.13$\\
  J1628+1154   & $0.97\pm0.06$ & ---\\
  J1659+2101   & $0.98\pm0.06$ & $1.19\pm0.10$\\
  J1720+3104   & $0.93\pm0.06$ & $0.97\pm0.08$\\
  J2102+6015   & --- & $1.51\pm0.13$ \& $1.24\pm0.09$\,$^{\mathrm{c, d}}$ \\ 
  J2228+0110   & --- & ---\\
  \hline
  \multicolumn{3}{p{\columnwidth}}{\footnotesize{\textbf{Columns:} Col.~1 -- source name (J2000); Col.~2 -- ratio of integrated FIRST to NVSS flux densities; Col.~3 -- ratio of `predicted' 1.4\,GHz VLBI flux density to FIRST flux density.}}\\
  \multicolumn{3}{p{\columnwidth}}{\footnotesize{\textbf{Notes:} $^{\mathrm a}$ The value was calculated using the flux densities of the 1.4\,GHz VLBI and VLA A configuration observations of the source reported in \citet{2005AJ....129.1809M}. $^{\mathrm b}$ The value was calculated using the \citet{2004AJ....127..587M} 1.4\,GHz VLBI flux density. $^{\mathrm c}$ The value was calculated using the NVSS flux density. $^{\mathrm d}$ The two values were calculated using the 2.3 to 8.3\,GHz and 2.3 to 8.6\,GHz spectral indices reported in Table~\ref{tbl:vlbi_z>4.5}, respectively.}}\\  
  \end{tabular}
  \label{tbl:Flux density ratios}
 \end{minipage}
\end{table}

Using a search radius of 2.5 and 23\,arcsec for FIRST and NVSS, respectively, we found unique matches for all of the sources except {\bf J0324$-$2918}, {\bf J0906+6930}, {\bf J1205$-$0742}, {\bf J1427+3312}, {\bf J2102+6015} and {\bf J2228+0110}. For J1205$-$0742, J1427+3312 and J2228+0110, the FIRST flux densities are all smaller than 1.32\,mJy. These sources are below the NVSS detection threshold of 2.5\,mJy\,beam$^{-1}$ \citep{nvss}. No match could be found for J0324$-$2918, J0906+6930 and J2102+6015 in FIRST because they lie outside the survey coverage. 

In Table~\ref{tbl:Flux density ratios}, the ratio of the sources' `predicted' 1.4\,GHz VLBI flux density and the FIRST flux density ($S_{\mathrm{pred}}/S_{\mathrm{FIRST}}$) is also given. As described in Section~\ref{subsec:Comments on Individual Sources}, $S_{\mathrm{pred}}$ is calculated for each source by extrapolating its measured VLBI flux density to 1.4\,GHz using its spectral index ($\alpha_{\mathrm{source}}$) and assuming a power-law radio spectrum. We consider $S_{\mathrm{pred}}$ as characteristic to the compact VLBI structure at 1.4\,GHz, which can be directly compared with the FIRST and NVSS values measured at this frequency. 

For assessing the variability, we first consider the $S_{\mathrm{FIRST}}/S_{\mathrm{NVSS}}$ values, defining a source to be variable if its FIRST and NVSS flux densities differ by more than 10\,per\,cent. If the flux density difference could be less than 10\,per\,cent inside the uncertainties we do not classify the source as being variable. As discussed later in this section, we note that this does not mean that the source is not variable. We also note that because of the difference in resolution between FIRST and NVSS, we only classify a source as being variable if the FIRST (the higher resolution catalogue) flux density is higher than the NVSS flux density. If the NVSS flux density is higher than the FIRST flux density, the difference could be because the source is resolved in FIRST, or it could be caused by variability. Based on their $S_{\mathrm{FIRST}}/S_{\mathrm{NVSS}}$ values, the only sources that could be variable are {\bf J0011+1446}, {\bf J0836+0054}, {\bf J1429+5447} and {\bf J1454+1109}.

{\bf J0011+1446} has a FIRST and NVSS flux density of $24.3\pm1.2$ and $35.8\pm1.5$\,mJy, respectively. Looking at the higher-resolution FIRST image, there is a second and third source that are 16.4 and 29.3\,arcsec away from J0011+1446 with flux densities of $3.0\pm0.2$ and $2.8\pm0.2$\,mJy, respectively. Since the NVSS beam size is 45\,arcsec \citep{nvss}, both of the nearby sources will blend with J0011+1446 in NVSS. However, the sum of the flux densities of these two sources is still less than the difference between the FIRST and NVSS flux densities. The remaining flux density difference could be explained by variability of J0011+1446 or of the two nearby sources. Given that we did not find evidence for variability in the multi-epoch EVN observations of J0011+1446, it is likely that J0011+1446 is non-variable. Another possible explanation is that J0011+1446 or the nearby sources are extended beyond 5\,arcsec and this structure may be resolved out in FIRST. From Table~\ref{tbl:Flux density ratios}, $S_{\mathrm{pred}}/S_{\mathrm{FIRST}}$ nearly equals 1 for J0011+1446. Considering that the source is not variable, this indicates that J0011+1446 is compact on angular scales between 5\,arcsec and $\sim5$\,mas. Hence, it is unlikely that J0011+1446 is extended beyond 5\,arcsec.

The FIRST and NVSS flux densities of {\bf J0836+0054} are $1.11\pm0.06$\,mJy and $2.5\pm0.5$\,mJy, respectively. This could indicate that J0836+0054 is extended at angular scales beyond 5\,arcsec. However, \citet{2003AJ....126...15P} observed J0836+0054 with the VLA in A configuration at 1.4\,GHz with a resolution of 1.5\,arcsec and found a flux density of $1.75\pm0.04$\,mJy. Considering that the \citet{2003AJ....126...15P} observations have higher resolution than FIRST and show $\sim60$\,per\,cent higher flux density, we conclude that J0836+0054 is variable. It appears that this conclusion is supported by $S_{\mathrm{pred}}$ being higher than that of FIRST (Table~\ref{tbl:Flux density ratios}). However, as no error is available for the predicted flux density, this cannot be said for certain.

The NVSS flux density of {\bf J1429+5447} ($3.8\pm0.5$\,mJy) is higher than that of FIRST ($3.0\pm0.2$\,mJy) and $S_{\mathrm{pred}}=4.0\pm0.2$\,mJy, which is equal to or slightly higher than the NVSS flux density. We therefore conclude that J1429+5447 is variable.

From Table~\ref{tbl:Flux density ratios}, the EVN flux density of {\bf J1454+1109} is higher than its FIRST flux density, which is higher than its NVSS flux density. We therefore conclude that J1454+1109 is variable. 

We now turn to $S_{\mathrm{pred}}/S_{\mathrm{FIRST}}$. {\bf J1146+4037}, {\bf J1427+3312}, {\bf J1429+5447}, {\bf J1454+1109}, {\bf J1606+3124}, {\bf J1611+0844} and {\bf J2102+6015} all have $S_{\mathrm{pred}}/S_{\mathrm{FIRST}}>1.1$. In the discussion above, we already concluded that J1429+5447 and J1454+1109 are variable. For J1146+4037, J1427+3312, J1606+3124, J1611+0844 and J2102+6015, $S_{\mathrm{pred}}/S_{\mathrm{FIRST}}>1.1$ could indicate that they are also variable. However, care should be taken. When calculating $S_{\mathrm{pred}}$ it is assumed that the spectral index of the source is constant between the two frequencies over which it is calculated. This assumption is not necessarily true since it is likely that at least some of the sources are gigahertz peaked-spectrum (GPS) and megahertz peaked-spectrum (MPS) sources \citep{coppejans2015,2016MNRAS.459.2455C}. GPS and MPS sources are radio-loud AGN that are identified based on their peaked spectra with steep optically thin spectra above the spectral turnover. The only difference between the GPS and MPS sources are that they have spectral turnovers above and below 1\,GHz, respectively\footnote{See \citet{o'dea1998} for a review of GPS sources.}. If a source has a spectral turnover between 1.7 and 5\,GHz, $S_{\mathrm{pred}}$ will be overestimated resulting in $S_{\mathrm{pred}}/S_{\mathrm{FIRST}}>1$, without the source being variable. In Section~\ref{sec:what are they?}, we argue that a significant fraction of the sources are GPS and MPS sources. Consequently, while it is possible that J1146+4037, J1427+3312, J1606+3124, J1611+0844 and J2102+6015 are variable based on their $S_{\mathrm{pred}}/S_{\mathrm{FIRST}}$ values, no definite statement can be made. No $S_{\mathrm{pred}}/S_{\mathrm{FIRST}}$ values are given for {\bf J0906+6930} and {\bf J1430+4204}, since the lowest of the frequencies used for determining their spectral index is above 5\,GHz. Consequently, there is a very large uncertainty of their predicted 1.4\,GHz flux density. Additionally, it is very likely that their spectra cannot be described by a simple power-law with a constant spectral index between 1.4 and 15\,GHz, and 1.4 and 43\,GHz, respectively. 

The most likely explanation for the sources with $S_{\mathrm{pred}}/S_{\mathrm{FIRST}}<1$ is that some of their radio emission is diffuse and extended beyond the angular scales probed by VLBI. Therefore they are partly resolved out. It is, however, also possible that they are variable. \citet{2001ApJ...555..625C} observed {\bf J1235$-$0003} at 1.4\,GHz with the VLA in A configuration and a resolution of $\sim1.5$\,arcsec, and found a flux density of $18.8\pm0.4$\,mJy. Considering that the NVSS and FIRST flux densities of J1235$-$0003 are $18.4\pm0.9$\,mJy and $18.1\pm0.7$\,mJy, respectively, and that its 1.4\,GHz VLBI flux density is within 7\,per\,cent of the FIRST flux density (Tables~\ref{tbl:vlbi_z>4.5} and \ref{tbl:Flux density ratios}), we conclude that J1235$-$0003 is likely not variable. In the case of {\bf J0913+5919}, comparing its 1.4\,GHz VLA flux densities with FIRST, \citet{2004AJ....127..587M} concluded that J0913+5919 is not variable. The flux density values of {\bf J0131$-$0321} in NVSS, FIRST, and the 1.7\,GHz VLBI flux density \citep{2015MNRAS.450L..57G} are $31.4\pm1.0$\,mJy, $33.7\pm1.7$\,mJy and $64.4\pm0.3$\,mJy, respectively. J0131$-$0321 was only observed at 1.7\,GHz with VLBI and we therefore do not have a spectral index for it. To reconcile the VLBI and FIRST flux densities, J0131$-$0321 would be required to have a spectral index of $\sim3.3$, which is unphysical. We therefore conclude that J0131$-$0321 is variable. {\bf J2228+0110} was also only observed at 1.7\,GHz with VLBI \citep{2014A&A...563A.111C}. Considering that its VLBI flux density in Table~\ref{tbl:vlbi_z>4.5} is significantly lower than its FIRST flux density of $1.32\pm0.07$\,mJy, it is likely resolved out.

{\bf J0311+0507}, {\bf J0913+5919}, {\bf J1235$-$0003}, {\bf J1242+5422}, {\bf J1659+2101} and {\bf J1720+3104} all have $0.9 < S_{\mathrm{FIRST}}/S_{\mathrm{NVSS}} <1.1$ and $0.9<S_{\mathrm{pred}}/S_{\mathrm{FIRST}} < 1.1$ within their uncertainties. This indicates that these sources are likely not variable, as was already concluded for J0913+5919 and J1235$-$0003 in the previous paragraph. In Section~\ref{subsec:multi epoch} we concluded from our multi-epoch EVN observations that J1720+3104 could be variable, apparently contradicting this result. Consequently, it is possible that J1720+3104 is variable. As mentioned in Section~\ref{subsec:multi epoch}, from the multi-epoch EVN observations of {\bf J0940+0526}, it is possible that it is variable.

It should be made clear that, just because its NVSS and FIRST flux densities are similar, an individual source cannot be claimed as non-variable. It is possible that an otherwise variable source was observed by NVSS and FIRST when it showed the same flux density by chance. In other words, regular flux density monitoring observations would be essential to reach a firm conclusion on the invariability of any individual radio source. This is clearly illustrated by {\bf J1430+4204} whose FIRST and NVSS flux densities are nearly identical (Table~\ref{tbl:Flux density ratios}) but for which \citet{1999MNRAS.308L...6F} and \citet{2010A&A...521A...6V} found strong 15\,GHz variability (Section \ref{subsec:litrature Comments on Individual Sources}). Similarly the two 2.3\,GHz VLBI observations of {\bf J2102+6015} (Table~\ref{tbl:vlbi_z>4.5}) indicate that it is not variable. However, to reconcile the 8.3 and 8.6\,GHz VLBI flux densities requires an unphysical spectral index of $\sim15$, indicating that J2102+6015 is variable. We note that in both cases, the lack of low frequency variability could be the result of the lower frequencies probing regions farther away form the black hole. In these regions, the variability amplitude will be lower, and the time scale will be longer than in the regions closer to the black hole, that are probed by the higher frequency observations.

In summary, it is striking that in only four of the 24 sources for which we have NVSS and FIRST flux densities, the values indicate variability, and even in one of those cases (J0011+1446) the difference is likely because of other factors. If the majority of the sources would be strongly variable, a few of them could have, by chance, been observed by FIRST and NVSS at the same flux density. However, the majority of the sample should still show variability. It therefore appears that most of the known $z>4.5$ VLBI radio sources are not significantly variable. This tendency is qualitatively consistent with the finding by \citet{2008ApJ...689..108L, 2011AJ....142..108K,2012ApJ...756...29K} that sources at $z\lesssim2$ are more variable than sources at $2\lesssim z\lesssim4$. We do, however, point out that the authors in these studies searched for variability on timescales of a few days and only studied flat-spectrum sources. Consequently, care should be taken when comparing the results. A dedicated follow-up study on the combined sample would allow conclusive statements to be made.

\subsection{Spectral index}
\label{subsec:alpha}
From the summary of the properties of the $z>4.5$ radio sources found in the literature (Table~\ref{tbl:vlbi_z>4.5}), the spectral indices of the sources with multiple components were calculated by summing the flux densities of all components at each frequency. {\bf J1548+3335} is the only source in our new EVN sample that is resolved into more than one component (Table~\ref{tbl:source parameters}). As noted in Section~\ref{subsec:Comments on Individual Sources}, its source spectral index equals $-1.07\pm0.07$. 

If we conventionally define a flat spectral index as $|\alpha_{\mathrm{source}}|<0.5$, there are nine sources that have flat spectra. Taking the uncertainties of the spectral index into consideration, there are three sources that have flat spectra, but could have positive or negative spectral indices within their uncertainties. Similarly, there are four sources that have negative spectral indices, but they could have flat spectra within the uncertainties.

The spectral index of all of the sources except {\bf J0906+6930} and {\bf J1430+4204} were calculated between observing frequencies of $\sim1.7$ and $\sim8$\,GHz. At $z=5$ this corresponds to the rest-frame frequencies of $\sim10$ and $\sim50$\,GHz, respectively. The spectral index for J1430+4204 was calculated between the rest-frame frequencies of 29 and 86\,GHz, while for J0906+6930 between 97 and 278\,GHz, assuming a power-law dependence of the flux density from the frequency. Excluding J1628+1154 (for which we only have a lower limit on its spectral index), J0906+6930 and J1430+4204, there are $8^{+4}_{-3}$ sources with flat spectra, out of the 22 objects which have reliable spectral indices. In other words, $36^{+18}_{-14}$\,per\,cent of the sources have flat VLBI source spectra between $\sim10$ and $\sim50$\,GHz in their rest frames. 

\subsection{Doppler boosting}
\label{subsec:Tb}
For the intrinsic brightness temperature of the sources we assume the equipartition brightness temperature of a relativistic compact jet. It is estimated as $T_{\mathrm{b,eq}} \simeq 5\times10^{10}$\,K \citep{1994ApJ...426...51R}. Since the Doppler factor is $\delta = T_{\mathrm{b}}/T_{\mathrm{b,eq}}$ \citep[e.g.][and references therein]{2010A&A...521A...6V}, the jet emission in sources with $T_{\mathrm{b}} > T_{\mathrm{b,eq}}$ is Doppler-boosted. We therefore conclude that 12 sources are Doppler-boosted.

Excluding the five sources for which we only have lower limits on $T_{\mathrm b}$ and that have $T_{\mathrm b} < T_{\mathrm{b,eq}}$, there are 13 sources out of the remaining 25 with $T_{\mathrm b} < T_{\mathrm{b,eq}}$. There are two possible physical reasons for why a source could have $T_{\mathrm{b}} < T_{\mathrm{b,eq}}$: (1) the jet viewing angle is moderate, resulting in the emission being Doppler-deboosted; (2) the flux density of the source is measured far away from the spectral peak frequency caused by synchrotron self-absorption. It is therefore possible that the sources with $T_{\mathrm b} < T_{\mathrm{b,eq}}$ could still be Doppler-boosted. Finally, brightness temperature is inversely proportional to the size of the source squared (Eq. \ref{eq:Tb}). A source could therefore have $T_{\mathrm{b}} < T_{\mathrm{b,eq}}$ because its size is overestimated or the error on its size is underestimated. In cases where the sources are too faint for phase self-calibration, or where the source components are not clearly resolved, the brightness temperatures should consequently be taken as lower limits.

\section{What are the z$>$4.5 VLBI sources?}
\label{sec:what are they?}
Based on their spectra, variability and whether or not they are Doppler-boosted, we now classify the $z>4.5$ radio sources studied with VLBI to date into three basic classes. A summary of the parameters used for the classification, and the classification of each source are given in Table~\ref{tbl:Source clasification}. 

The first one, which only contains a single object, {\bf J1205$-$0742}, is the class of {\it star-forming sources}.
The second class is the {\it flat-spectrum radio quasars} (FSRQs). {\bf J1454+1109} is the archetypal example of a FSRQ, having a flat spectrum over a wide spectral range, showing variability and being Doppler-boosted. {\bf J1430+4204} similarly satisfies all three criteria, albeit the frequencies used for calculating its spectral index are higher than for the other sources. We do, however, note that calculating its spectral index based on the integrated 5\,GHz flux density and the 2.3\,GHz peak intensity gives $\alpha_{\mathrm{source}} =-0.38$, which also indicates a flat spectrum. While there is no solid evidence that {\bf J0940+0526}, {\bf J1606+3124} and {\bf J1720+3104} are variable, there are indications that they are (Section \ref{subsec:variability}), and they all have flat radio spectra and Doppler-boosted jet emission. {\bf J0210$-$0018}, {\bf J0324$-$2918} and {\bf J1013+2811} are FSRQs since they have flat spectra and are Doppler-boosted. {\bf J0131$-$0321}, {\bf J0913+5919} and {\bf J1026+2542} join the FSRQs based on Doppler-boosting, and, in the case of {\bf J0131$-$0312}, variability. For the three sources above, we do not have spectral information regarding their VLBI structure. In Section~\ref{subsec:variability} we concluded that J0913+5919 is not variable. While this seems to indicate that it is not a FSRQ, non-variability based on a few sparsely obtained flux density measurements cannot be regarded as certain. Therefore we leave it in this class, based on its high measured brightness temperature as the evidence for Doppler-boosting. J1026+2542 has a steep negative spectral index, which would also contradict the FSRQ classification. However, this is likely the result of two components which are resolved in the 5\,GHz VLBI observations but not at 1.7\,GHz (Table~\ref{tbl:vlbi_z>4.5}). The steep spectrum could also (at least in part) be caused by variability. {\bf J2102+6015} is a FSRQ because it is Doppler-boosted and variable, the latter of which likely explains its steep negative spectral index. Finally {\bf J1611+0844} joins the FSRQs based on its flat spectrum and indications of variability.

\begin{table*}
 \hspace{4cm}
 \centering
 \begin{minipage}{\textwidth}
  \caption{Classification of the $z>4.5$ sources}
  \begin{tabular}{cccccc}
  \hline
  ID       & Origin of the radio emission$^{\mathrm{e}}$ & Spectrum$^{\mathrm{a,c}}$ & Variable$^{\mathrm c}$ & Boosted$^{\mathrm c}$ & Classification$^{\mathrm c}$ \\
  (1)      & (2)                          & (3)                       & (4)                    & (5)                   & (6) \\
  \hline  
  J0011+1446   & AGN: $T_{\mathrm b}$ \& $L$   & Negative (flat)        &          &     & Steep-spectrum \\
  J0131$-$0321 & AGN: $T_{\mathrm b}$          &                        & Yes      & Yes & FSRQ \\
  J0210$-$0018 & AGN: $T_{\mathrm b}$ \& $L$   & Flat (positive)        &          & Yes & FSRQ \\
  J0311+0507   & AGN: $T_{\mathrm b}$ \& $L$   & Negative               & No       &     & Steep-spectrum/wide double \\
  J0324$-$2918 & AGN: $T_{\mathrm b}$ \& $L$   & Flat (Negative)        &          & Yes & FSRQ \\
  J0813+3508   & AGN: $T_{\mathrm b}$ \& $L$   & Negative               &          &     & Steep-spectrum \\
  J0836+0054   & AGN: $T_{\mathrm b}$ \& $L$   & Negative               & Yes      &     & Steep-spectrum \\
  J0906+6930   & AGN: $T_{\mathrm b}$ \& $L$   & Negative$^{\mathrm b}$ &          &     & Steep-spectrum \\
  J0913+5919   & AGN: $T_{\mathrm b}$          &                        & No       & Yes & FSRQ \\
  J0940+0526   & AGN: $T_{\mathrm b}$ \& $L$   & Flat                   & Possibly & Yes & FSRQ \\
  J1013+2811   & AGN: $T_{\mathrm b}$ \& $L$   & Flat (Negative)        &          & Yes & FSRQ \\
  J1026+2542   & AGN: $T_{\mathrm b}$ \& $L$   & Negative               &          & Yes & FSRQ \\
  J1146+4037   & AGN: $T_{\mathrm b}$ \& $L$   & Negative (flat)        & Possibly &     & Steep-spectrum \\
  J1205$-$0742 & Star-forming\,$^{\mathrm{d}}$ &                        &          & No  & Star-forming \\
  J1235$-$0003 & AGN: $T_{\mathrm b}$          &                        & No       &     &  \\
  J1242+5422   & AGN: $T_{\mathrm b}$ \& $L$   & Negative (flat)        & No       &     & Steep-spectrum \\
  J1311+2227   & AGN: $T_{\mathrm b}$ \& $L$   & Negative               &          &     & Steep-spectrum \\
  J1400+3149   & AGN: $T_{\mathrm b}$ \& $L$   & Negative (flat)        &          &     & Steep spectrum \\
  J1427+3312   & AGN: $T_{\mathrm b}$ \& $L$   & Negative               & Possibly &     & Steep-spectrum \\
  J1429+5447   & AGN: $T_{\mathrm b}$ \& $L$   & Negative               & Yes      &     & Steep-spectrum \\
  J1430+4204   & AGN: $T_{\mathrm b}$ \& $L$   & Flat$^{\mathrm b}$     & Yes      & Yes & FSRQ \\
  J1454+1109   & AGN: $T_{\mathrm b}$ \& $L$   & Flat                   & Yes      & Yes & FSRQ \\
  J1548+3335   & AGN: $T_{\mathrm b}$ \& $L$   & Negative               &          &     & Steep-spectrum/wide double\\
  J1606+3124   & AGN: $T_{\mathrm b}$ \& $L$   & Flat                   & Possibly & Yes & FSRQ \\
  J1611+0844   & AGN: $T_{\mathrm b}$ \& $L$   & Flat                   & Possibly &     & FSRQ \\
  J1628+1154   & AGN: $T_{\mathrm b}$          &                        &          &     &  \\
  J1659+2101   & AGN: $T_{\mathrm b}$ \& $L$   & Negative               & No       &     & Steep-spectrum \\
  J1720+3104   & AGN: $T_{\mathrm b}$ \& $L$   & Flat                   & Possibly & Yes & FSRQ \\ 
  J2102+6015   & AGN: $T_{\mathrm b}$ \& $L$   & Negative               & Yes      & Yes & FSRQ \\
  J2228+0110   & AGN: $T_{\mathrm b}$          &                        &          &     &  \\
  \hline
  \multicolumn{6}{p{15cm}}{\footnotesize{\textbf{Columns:} Col.~1 -- source name (J2000); Col.~2 -- the origin of the radio emission from Section~\ref{subsec:agn star}; Col.~3 -- spectral classification from Section~\ref{subsec:alpha}; Col.~4 -- source variability from Section~\ref{subsec:variability}; Col.~5 -- Doppler boosting from Section \ref{subsec:Tb}; Col.~6 -- source classification.}}\\
  \multicolumn{6}{p{15cm}}{\footnotesize{\textbf{Notes:} $^{\mathrm a}$ Wording such as `Negative (flat)' indicates that the source has a negative spectral index, but that it could be flat within the uncertainties. $^{\mathrm b}$ The spectral index is calculated between higher rest frame frequencies than for the other sources. See Section~\ref{subsec:alpha}. $^{\mathrm c}$ Blank space indicates that the property could not be determined because of insufficient information. $^{\mathrm d}$ See Section \ref{subsec:litrature Comments on Individual Sources}. $^{\mathrm e}$  Wording such as `AGN: $T_{\mathrm b}$ \& $L$' indicates that the source is AGN powered based on both its brightness temperature and 1.4\,GHz luminosity.}}\\  
  \end{tabular}
  \label{tbl:Source clasification}
 \end{minipage}
\end{table*}

The final, third class of objects is the {\it steep-spectrum sources}. All of them have negative spectral indices with $\alpha_{\mathrm{source}} < -0.5$, are therefore not FSRQs, and have $T_{\mathrm{b}} < T_{\mathrm{b,eq}}$. Since their flux densities cannot continue to increase indefinitely toward lower frequencies, their spectra will turn over because of synchrony self absorption at some point. These sources are thus likely MPS or GPS sources, something that could be proven using low-frequency ($<1$\,GHz) observations. We note that the two wide-separation double sources ({\bf J0311+0507} and {\bf J1548+3335}) are also included in this class. For the majority of the sources in this class, there is either insufficient information about variability, or we concluded that they are not variable in Section~\ref{subsec:variability}. The exceptions to this are {\bf J0836+0054}, {\bf J1146+4037}, {\bf J1427+3312} and {\bf J1429+5447} for which we found indication of variability or concluded that they are variable in Section~\ref{subsec:variability}. While the GPS (and therefore the MPS) sources are the least variable class of radio sources \citep{o'dea1998}, there are GPS sources which show significant variability \citep[e.g.][]{1992ApJ...391..589W,o'dea1998}. Hence, variability does not necessarily mean that the source is not a GPS or MPS source. We also note that some of the flat-spectrum sources could be GPS sources with a spectral turnover near the center of the frequency range over which the spectral index is measured. Finally, for {\bf J1235$-$0003}, {\bf J1628+1154} and {\bf J2228+0110}, the information available at present is not sufficient to assign them with any of the three classes above.    

\subsection{Correlations between variability, brightness temperature and spectral index}
\label{subsection: Correlations with variablity}
To check for potential correlations between variability and brightness temperature/spectral index, we derived a characteristic brightness temperature and spectral index for each of the sources using the values in Tables~\ref{tbl:source parameters} and \ref{tbl:vlbi_z>4.5}. For both spectral index and brightness temperature, the characteristic value is the average of all of the values for the source that are not upper or lower limits. The median spectral index and brightness temperature of the sources that are indicated as being variable, or possibly variable, in Table~\ref{tbl:Source clasification} are $-0.38$ and $1.5\times10^{10}$\,K, respectively. Repeating this for the remaining sources, gives median values of $-0.65$ and $3.9\times10^{9}$\,K for the spectral index and brightness temperature, respectively. 

We note that caution should be taken when interpreting these values since there is insufficient information to say whether 13 of the sources are variable or not and, as mentioned previously, regular flux density monitoring observations are essential to reach a firm conclusion on the invariability of any individual radio source. Despite this, it would appear that the sources in our sample that are variable, or for which there are indications that they are variable, have flatter spectra and higher brightness temperatures than the other sources. This result fits the picture that the variable sources are FSRQs, which have flat spectra and in which the emission is Doppler boosted, which causes their brightness temperatures to be higher than for the other classes. 

We finally also checked for a correlation between brightness temperature and spectral index, but could not find one.

\section{Summary and conclusions}
\label{sec:summary}
In this paper, we presented 1.7 and 5\,GHz EVN observations of ten $z>4.5$ sources. These observations increased the number of $z>4.5$ sources that have been imaged with VLBI by 50\,per\,cent from 20 to 30. Combining our new sources with those from the literature, we investigated the origin of the radio emission, variability properties, spectral indices and Doppler boosting of the $z>4.5$ VLBI sources (Section \ref{sec:discus}). Based on these properties we classified the sources as star-forming, flat-spectrum radio quasars (FSRQ) and steep-spectrum sources (Section \ref{sec:what are they?}).

Of the 27 sources that can be classified, 13 (48\,per\,cent) are FSRQs. One of the sources is star-forming, illustrating that even at $z>4.5$ not all VLBI-detected sources are necessarily AGN. The remaining 13 objects are steep-spectrum sources that are likely GPS or MPS sources. The idea that a large fraction of the high-redshift sources are GPS and MPS sources was first proposed by \citet{1982ApJ...260L..27P} and \citet{1991ApJ...380...66O}. Our conclusion that $\sim50$\,per\,cent of the sources are GPS and MPS sources is supported by the finding in Section~\ref{subsec:variability} that the majority of the sources are not significantly variable, since the GPS and MPS sources are the least variable class of radio sources \citep{o'dea1998}. It is likely that our classification is uncertain for a few sources, i.e. some of the steep-spectrum sources are in fact FSRQs and vice versa. Despite this, it is clear that roughly half of the sources are FSRQs. This result seems to support the finding by \citet{2011MNRAS.416..216V} that beyond $z=3$, the number of high-redshift radio-loud sources is significantly lower than what is expected from the number of blazars at these redshifts (Section \ref{sec:introduction}). However, it is difficult to accurately assess the selection effects that are at play here. The presently known sample of $z>4.5$ radio sources imaged with VLBI is incomplete and rather inhomogeneous, but can be regarded as optically selected because of their measured spectroscopic redshifts. Moreover, like in the case of the new 10-element sample reported in this paper, the selection for follow-up high-resolution VLBI observations generally involves a flux density lower limit and arcsec-scale structural compactness based on e.g. FIRST data. Both the optical and radio selection criteria may result in a sample biased towards sources with high luminosities, high radio-loudness and Doppler-boosted radio emission. In this respect, not the significant fraction of FSRQs in Table~\ref{tbl:Source clasification} is surprising, but the relatively large number of steep-spectrum sources with unboosted radio jet emission.

While the new observations presented in this paper significantly increased the number of $z>4.5$ sources imaged with VLBI, the overall number of sources that have been observed is still very small. Continuing observing efforts are therefore needed to classify new sources as star-forming, FSRQ or steep-spectrum, address open questions such as whether the average bulk Lorentz factor evolves with redshift and to test cosmological models. Complementary to this, non-VLBI observations at frequencies between 100\,MHz and $\sim20$\,GHz will allow the sources to be classified based on their broad-band spectra and variability. Specifically, observations below 1.4\,GHz will allow the GPS and MPS nature of the steep-spectrum sources to be confirmed or refuted.

\section*{Acknowledgements}
The authors wish to thank the anonymous referee for their suggestions and comments which helped to improve this paper.
The European VLBI Network is a joint facility of independent European, African, Asian, and North American radio astronomy institutes. Scientific results from data presented in this publication are derived from the following EVN project code: EC052 (PI: D. Cseh).
S.F., D.C., and K.\'E.G. thank the Hungarian National Research, Development and Innovation Office (OTKA NN110333) for their support. 
This work was partly supported by the China--Hungary Collaboration and Exchange Programme by the International Cooperation Bureau of the Chinese Academy of Sciences.
C.M. and H.F. is funded by the ERC Synergy Grant BlackHoleCam: Imaging the Event Horizon of Black Holes (Grant 610058).


\bibliographystyle{mnras.bst}
\bibliography{references.bib}

\label{lastpage}

\end{document}